\documentclass[pdflatex,sn-mathphys-num]{sn-jnl}

\usepackage{graphicx}%
\usepackage{multirow}%
\usepackage{amsmath,amssymb,amsfonts}%
\usepackage{amsthm}%
\usepackage{mathrsfs}%
\usepackage[title]{appendix}%
\usepackage{xcolor}%
\usepackage{textcomp}%
\usepackage{manyfoot}%
\usepackage{booktabs}%
\usepackage{physics}
\usepackage{algorithm}%
\usepackage{algorithmicx}%
\usepackage{algpseudocode}%
\usepackage{listings}%
\usepackage{gensymb}
\usepackage[separate-uncertainty=true,per-mode=fraction]{siunitx}

\theoremstyle{thmstyleone}%
\theoremstyle{thmstyletwo}%
\theoremstyle{thmstylethree}%

\begin{document}

\title[Scalable Quantum Photonic Platform Based on Site-Controlled Quantum Dots Coupled to Circular Bragg Grating Resonators]{Scalable Quantum Photonic Platform Based on Site-Controlled Quantum Dots Coupled to Circular Bragg Grating Resonators}


\author*[1]{\fnm{Kartik} \sur{Gaur}}\email{kartik.gaur@tu-berlin.de}
\author[1]{\fnm{Avijit} \sur{Barua}}
\author[1]{\fnm{Sarthak} \sur{Tripathi}}
\author[1]{\fnm{Léo J.} \sur{Roche}}
\author[2]{\fnm{Steffen} \sur{Wilksen}}
\author[2]{\fnm{Alexander} \sur{Steinhoff}}
\author[1]{\fnm{Sam} \sur{Baraz}}
\author[1]{\fnm{Neha} \sur{Nitin}}
\author[1]{\fnm{Chirag C.} \sur{Palekar}}
\author[1]{\fnm{Aris} \sur{Koulas-Simos}}
\author[1]{\fnm{Imad} \sur{Limame}}
\author[1]{\fnm{Priyabrata} \sur{Mudi}}
\author[1]{\fnm{Sven} \sur{Rodt}}
\author[2]{\fnm{Christopher} \sur{Gies}}
\author*[1]{\fnm{Stephan} \sur{Reitzenstein}}\email{stephan.reitzenstein@physik.tu-berlin.de}

\affil[1]{\orgdiv{Institut für Physik und Astronomie}, \orgname{Technische Universität Berlin}, \orgaddress{\street{ Hardenbergstraße 36}, \city{Berlin}, \postcode{10623}, \country{Germany}}}

\affil[2]{\orgdiv{Institute for Physics}, \orgname{Carl von Ossietzky Universität Oldenburg}, \orgaddress{\city{Oldenburg}, \postcode{26129}, \country{Germany}}}


\abstract{The scalable integration of solid-state quantum emitters into photonic nanostructures remains a central challenge for quantum photonic technologies. Here, we demonstrate a robust and streamlined integration strategy that tackles the long-standing issue of deterministic fabrication on randomly positioned self-assembled quantum dots (QDs), leveraging a buried-stressor-based site-controlled InGaAs QD platform. We show that this deterministic growth approach enables precise spatial alignment with circular Bragg grating (CBG) resonators for enhanced emission, eliminating the need for complex and time-consuming deterministic lithography techniques. We fabricated a $6\times6$ SCQD-CBG array with 100\% device yield, with 35 devices falling within the radial-offset range where the simulated photon-extraction efficiency (PEE) exceeds 20\%, underscoring the spatial precision and scalability of our fabrication concept. A systematically selected subset of five devices with varying radial displacements reveals clear offset-dependent trends in extraction efficiency, spectral linewidth, and photon indistinguishability, thereby establishing quantitative bounds on spatial alignment tolerances. In the best-aligned QD-CBG device, we achieve a PEE of $(47.1\pm3.8)\%$, a linewidth of $(1.41\pm0.22)$ GHz, a radiative decay lifetime of $(0.80\pm0.02)$~ns, a single-photon purity of $(99.58\pm0.18)\%$, and a Hong-Ou-Mandel two-photon interference visibility of $(81\pm5)\%$ under quasi-resonant excitation at saturation power. We confirm our conceptual understanding of the effect of emitter-position dependent charge-noise fluctuations in terms of a quantum-optical model for the (quantum-)emission properties. The established nanofabrication platform provides a reproducible, lithography-compatible route to scalable, high-performance single-photon sources (SPS), offering a powerful alternative to conventional lithography-based deterministic integration techniques.}




\maketitle

\section{Introduction}\label{sec1}
Semiconductor quantum dots (QDs) constitute one of the most powerful solid-state platforms for generating non-classical light, combining near-unity internal quantum efficiency, discrete energy levels, and compatibility with integrated photonic architectures~\cite{Michler2000quantum, Senellart2017, Arakawa2020, Heindel2023, Rodt2024, Maring2024}. Their potential for deterministic single-photon emission, photon indistinguishability, and electrical operability has positioned them at the forefront of photonic quantum information technology~\cite{Heindel2023}, enabling proof-of-principle demonstrations of boson sampling~\cite{Wang2017, Loredo2017}, quantum teleportation~\cite{Fattal2004, Pirandola2015, BassoBasset2021}, photonic quantum computing~\cite{Ladd2010, Maring2024, aghaeerad2025}, and elementary quantum networks~\cite{Wang2020, Lu2021, Yu2023, Strobel24}. Embedding QDs into optical microcavities, waveguides, or metastructures enables enhanced light-matter interaction, improving brightness, directionality, and coherence of the emitted photons~\cite{Reitzenstein2012_IEEE, Kolatschek2021, Chanana2022}. However, achieving simultaneously high photon extraction efficiency (PEE) and near-unity indistinguishability remains nontrivial. Although a reduced radiative lifetime resulting from genuine Purcell enhancement can shorten the dephasing window and improve HOM visibility~\cite{Rickert2025}, practical approaches to increasing brightness, such as stronger optical confinement or higher excitation powers, often introduce charge noise, spectral diffusion, and phonon-assisted processes that broaden the emission linewidth and reduce photon indistinguishability.~\cite{Weiler2012, Ding2016}. Careful optimization of the emitter environment and surrounding photonic structure, combined with resonant or quasi-resonant excitation schemes, can mitigate these effects and preserve coherence~\cite{Somaschi2016, Uppu2020, Zhai2022}. In this context, scalable arrays of QD-based single-photon sources (SPS) are highly desirable for large-scale quantum networks and photonic processors, where multiple indistinguishable emitters operating in parallel can enable on-chip entanglement distribution, multiplexed photon generation, and high-throughput quantum information processing~\cite{Lodahl2015, Wang2025}. However, the intrinsic randomness in the spatial and spectral positioning of self-assembled QDs fundamentally impedes large-scale integration~\cite{Norman2018, Hepp2019}. Realizing spatial overlap between cavity modes and quantum emitters typically requires pre-characterization through cathodoluminescence (CL)~\cite{Shulun2023, madigawaDonges2024} or photoluminescence (PL) mapping~\cite{Dousse2008, Sapienza2015, Liu2017, madigawaDonges2024}, followed by marker-based or in-situ lithography~\cite{Gschrey2013, Gschrey2015, Rodt2021}, processing steps that introduce complexity, reduce throughput, and challenge scalability. In parallel, hybrid integration strategies based on transfer printing or "pick-and-place" assembly have also been explored, where pre-fabricated QDs or nanomembrane cavities are physically transferred onto target photonic circuits, but these approaches remain experimentally demanding and inherently limited in scalability~\cite{Zadeh2016, Katsumi2018}. Moreover, such approaches still rely on random QD positions, demand high-resolution imaging infrastructure and tight process control, restricting accessibility to well-equipped fabrication environments.

Site-controlled quantum dot (SCQD) platforms provide a pathway to overcome this barrier by enabling deterministic QD nucleation at lithographically defined locations~\cite{Gaur2025MQT, Hou2025}. Several techniques, such as surface patterning~\cite{Ishikawa2000, Baier2004, Schneider2008, Felici2009, cheng2009} and selective area epitaxy~\cite{Birudavolu2004, Mokkapati2005}, have been explored to engineer QD placement with nanometer-scale precision. While these SCQD approaches have been used to demonstrate initial attempts at scalable integration into nanophotonic resonators~\cite{Schneider2008, Sunner08, Schneider2009}, their optical performance remained suboptimal, preventing them from reaching the benchmarks needed for scalable quantum-photonic systems. Beyond these, the buried-stressor approach has emerged as a promising candidate, leveraging strain fields from subsurface oxide apertures to guide QD formation at the center of patterned mesas~\cite{StrittmatterPSSA, StrittmatterAPL, Gaur2025MQT}. This technique integrates seamlessly with standard planar growth protocols and, unlike surface-etched templates, avoids detrimental interface states that can compromise optical performance~\cite{Albert2010}. Furthermore, the stressor aperture size can be fine-tuned via oxidation control~\cite{ImadAPL, Podhorsky2024}, enabling systematic control over local QD density and site precision, which enables the fabrication of both quantum light sources~\cite{JanAPLPhotonics2020} and high-$\beta$ microlasers~\cite{ArsentyOptica} based on SCQDs. While initial realizations of buried-stressor SCQDs faced challenges related to linewidth broadening and low radiative efficiency, recent progress in epitaxial optimization has led to substantial improvements in emitter quality, including reduced spectral diffusion and enhanced coherence, opening the door to their use in quantum photonic devices~\cite{Unrau2012, MaxStrauss2017, JanAPLPhotonics2020, ArsentyElsevier}.

Despite this, the integration of SCQDs into high-performance photonic structures has remained largely reliant on some form of emitter pre-localization or alignment~\cite{ArsentyElsevier}. In particular, achieving consistent spatial overlap between the quantum emitter and the cavity mode without reverting to feedback-based positioning has proven nontrivial. This challenge is especially pronounced in nanophotonic cavities such as circular Bragg gratings (CBGs), where even slight displacements of the emitter from the center of the cavity can markedly reduce PEE and can induce linear polarization of the emitted photons~\cite{Rickert2019, Shih2023, Peniakov2024}. Moreover, scalable integration strategies must remain compatible with standard and ideally industry-relevant nanofabrication workflows, minimize reliance on complex lithographic procedures, and ensure high optical quality across large device arrays. Bridging this gap requires an approach that combines deterministic emitter positioning with a fabrication process that is alignment-free, spectrally flexible, and scalable beyond a few isolated devices.

Here, we overcome this integration challenge by developing a marker-free, lithography-compatible nanofabrication platform that exploits the intrinsic spatial precision of buried-stressor-grown SCQDs for direct integration into CBG resonators. In this approach, deterministic QD nucleation at the center of the site-controlled growth (SCG) mesas enables device fabrication entirely through standard electron beam lithography (EBL), eliminating the need for emitter localization via CL or PL mapping or in-situ EBL typically required for randomly positioned QDs. CBGs are patterned directly on the predefined mesa grid without optical mapping or reference markers, streamlining the process and enhancing device yield. The workflow operates entirely at room temperature and remains fully compatible with planar nanofabrication, circumventing any low-temperature alignment or processing steps. By decoupling cavity integration from emitter localization, it establishes a scalable, reproducible, and fabrication-efficient route toward wafer-level implementation of bright SPS with high photon indistinguishability.

\section{Results and discussion}\label{sec2}
In the following subsections, we introduce and discuss our scalable buried-stressor-based quantum device fabrication concept, presenting the epitaxial growth of SCQDs, the design and nanofabrication of SCQD-CBGs, the demonstration of scalable integration, and the systematic optical characterization of the resulting devices in combination with theoretical modeling of the quantum-emission properties.
\begin{figure}
 \centering
  \includegraphics[width=1.0\textwidth]{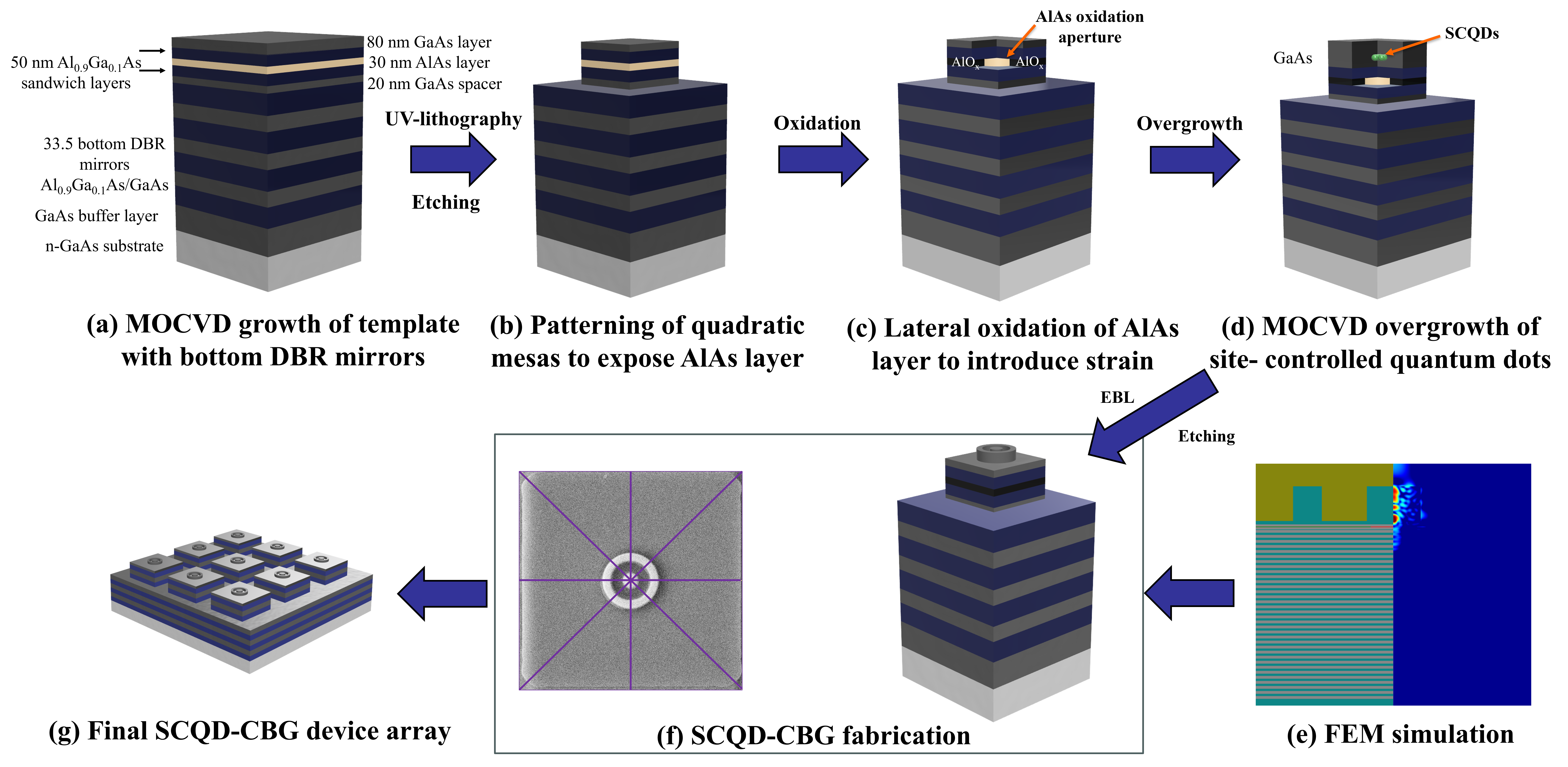}
  \caption{Schematic process flow for SCQD-CBG device fabrication. The fabrication pathway begins with MOCVD growth of a DBR template incorporating a buried AlAs stressor layer (a), followed by SCG-mesa definition through UV-lithography and etching (b) and lateral oxidation to form apertures at the center of the mesas (c). SCQDs are subsequently overgrown within a GaAs cavity by MOCVD (d). Prior to patterning, the CBG parameters were optimized through FEM simulations, as illustrated by the schematic device representation and the corresponding simulated electric field distribution (e). The CBGs with numerically optimized geometry are subsequently patterned on the SCG-mesas using EBL and etching (f), also shown in the representative SEM image of a completed SCQD-CBG device. This process yields an array of SCQD-CBG structures (g), ready for optical characterization.
}
  \label{fig:1}
\end{figure}

\subsection{Site-controlled quantum dot growth via the buried-stressor approach}\label{subsec1}
The growth of InGaAs SCQDs via the buried-stressor approach begins with metal-organic chemical vapor deposition (MOCVD) growth at 700\degree C of a 300~nm GaAs buffer layer on an n-type GaAs (001) substrate, followed by 33.5 Al$_{0.9}$Ga$_{0.1}$As/GaAs DBR mirror pairs and a 20~nm GaAs barrier layer. The buried-stressor layer comprises a 30~nm AlAs core encapsulated between two 50~nm Al$_{0.9}$Ga$_{0.1}$As layers, capped by 80~nm of GaAs. This epitaxial template is lithographically patterned into $\sim$20~$\mu$m square SCG-mesas to expose the AlAs layer laterally and is subsequently transferred to a controlled wet thermal oxidation process known from standard industry-compatible VCSEL fabrication. Conducted at 420\degree C in a water vapor/N$_2$ atmosphere, the selective lateral oxidation of the AlAs layer converts its periphery into AlO$_x$, leaving a central unoxidized AlAs aperture. The final aperture diameter is precisely defined by the lateral oxidation length, which is controlled via the SCG-mesa size in 100~nm increments, enabling precise tuning of the resulting tensile strain field. Real-time optical contrast monitoring ensures controlled termination of the oxidation to yield aperture openings between a few hundred nanometers and $\sim$2~$\mu$m. To remove native oxides, the sample undergoes a surface treatment consisting of a 30~s dip in 75\% H$_2$SO$_4$, followed by thorough rinsing in a deionized-water cascade until all acid residues are eliminated. The cleaned sample is then dried and immediately loaded into the MOCVD reactor for the subsequent overgrowth step. A 50~nm GaAs overgrowth layer is deposited to planarize the surface, followed by the growth of ultralow-density (1*10$^8$/cm$^2$) In$_{0.33}$Ga$_{0.67}$As QDs at 500\degree C, selectively nucleated at the center of the positions of maximum tensile strain~\cite{StrittmatterAPL, ImadAPL}. A thin 0.7~nm GaAs capping layer, also grown at 500\degree C, and a final 960~nm overgrowth layer, grown at 615\degree C, complete the epitaxial structure. The complete fabrication workflow for realizing SCQD-CBG devices is illustrated in Fig. \ref{fig:1}, outlining the sequential steps from DBR template growth and SCG-mesa patterning to SCQD overgrowth and final CBG patterning (discussed in subsection \ref{subsec4}), ultimately yielding a fully integrated device array.

\subsection{Optical assessment of SCG-mesas}\label{subsec2}
The optical and spatial uniformity of the SCG-mesas was systematically evaluated prior to photonic cavity integration. Micro-photoluminescence ($\mu$PL) spectroscopy was performed on over 200 individual SCG-mesas across a $5\times5$ mm$^2$ sample piece to assess emission wavelength homogeneity and verify emitter yield. A detailed description of the experimental setup is provided in the Supplementary Information (SI) Section 1 (Fig. S 1). The measurements reveal consistent spectral characteristics across the device array, with the dominant excitonic transitions distributed between 925~nm and 945~nm. This wavelength spread reflects stable strain and composition control during the buried-stressor growth process. The absence of any pronounced spectral outliers underscores the uniformity of the growth environment across the mesoscale array and confirms the reproducibility of the site-selective nucleation process. Exemplary emission spectra of SCG-mesas, together with a statistical histogram illustrating the emission wavelength distribution of over 200 individual structures across the sample, are provided in Fig. S 2 of the SI.

\subsection{Design and simulation of CBG resonators}\label{subsec3}


The design parameters for the CBGs were numerically optimized by calculating electric field distributions and maximizing the PEE into a collection numerical aperture (NA) of 0.81, matching the objective lens used in our optical experiments. We performed simulations using the finite-element method (FEM) scattering solver in JCMsuite~\cite{JCMsuite}, employing cylindrical rotational symmetry to retain full electromagnetic accuracy while minimizing computational costs. A Bayesian optimization algorithm was employed to simultaneously explore multiple structural degrees of freedom, including CBG-mesa diameter, ring thickness, ring gap, etching depth, and GaAs overgrowth thickness above the QD, each treated as a free parameter within fabrication-accessible bounds~\cite{Pomplun2007} (refer to SI Table S 1). Recent analysis has shown that even a single-ring CBG can deliver PEEs comparable to multi-ring architectures, while providing a substantially smaller device footprint and reduced fabrication complexity, which motivates our adoption of a single-ring design~\cite{barua2025}. The SCQD was modeled as a point dipole emitter placed 135~nm above the buried-stressor layer, a height chosen to reflect the strain-driven vertical positioning of our SCQD growth. The oxidation aperture was fixed at 800~nm in diameter~\cite{ArsentyElsevier}, aligned with experimentally achievable values~\cite{ArsentyOptica}. Among the optimized parameters, the GaAs overgrowth thickness plays a particularly critical role, influencing both the vertical cavity mode profile and practical growth feasibility. The algorithm autonomously converged toward geometries that ensured a maximal upward radiation into the desired NA cone and provided a direct blueprint for subsequent epitaxial and lithographic processes. The resulting optimized electric field distribution is shown in Fig. \ref{fig:1} (e) (see SI Section 3 for the high-resolution image).

To evaluate the robustness of this resonator design under realistic experimental conditions, we further performed scattering-mode simulations for PEE calculation in JCMsuite to quantify the impact of two key non-idealities: spectral mismatch and spatial misalignment of the emitter relative to the CBG resonator. In the first scenario, the emission wavelength of the dipole was swept from 900 to 960~nm, revealing the bandwidth tolerance of the optimized CBG and giving its compatibility with the natural inhomogeneity of QDs (see SI Fig. S 3). In the second scenario, the dipole position was systematically offset laterally from the CBG-mesa center, reflecting the expected variability in QD nucleation even under site-controlled conditions. The quantitative implications of emitter-CBG misalignment are analyzed in detail in the subsequent section \ref{subsubsec6}, where simulated offset-dependent PEEs are directly correlated with experimental measurements. 

\subsection{Marker-free scalable integration of SCQDs into CBGs}\label{subsec4}
The integration of SCQDs into CBG resonators was realized through EBL using a Raith eLINE Plus system, employing a marker-free process that fully relies on the spatial determinism intrinsic to the buried-stressor growth technique. Unlike deterministic QD-cavity integration approaches that depend on resource-intensive imaging modalities such as CL or PL mapping with respect to alignment markers, or on in-situ EBL, our method obviates the need for such procedures entirely. Notably, it also avoids the need for repeated cryogenic cooling cycles typically required for emitter localization, further simplifying the overall process flow. A $5\times5$ mm$^2$ sample containing a $16\times15$ array of SCG-mesas was fabricated, from which a subset of $6\times6$ SCG-mesas was selected for CBG integration. Owing to the strain-induced control of QD nucleation, achieved through precise oxidation kinetics and strain localization, the QDs consistently form within $\sim$100~nm-scale proximity to the SCG-mesa's center~\cite{StrittmatterAPL}. Consequently, each CBG resonator was patterned without performing any individual alignment of the structure to the center of the SCG-mesa. Instead, the EBL process was initiated by navigating to a known SCG-mesa from the mask layout, performing standard write-field alignment and beam optimizations, and subsequently executing a matrix-based exposure using a predefined pitch parameter of 368~$\mu$m between each mesa center, thus enabling automated and reproducible patterning across the full SCQD array. The scanning electron microscope (SEM) image in Fig. \ref{fig:1} (f) showcases the remarkably high precision of the fabrication process. Not only does our process simplify the workflow, but it also eliminates the need for alignment markers and auxiliary fabrication steps such as marker patterning via EBL and subsequent gold deposition. This streamlined fabrication flow, fully compatible with standard lithography techniques, enables scalable device realization across large-area samples. As such, the deterministic positioning of SCQDs and the simplicity of the integration protocol establish this approach as a compelling alternative to advanced image-based alignment techniques, offering a viable path for high-yield quantum photonic device fabrication even in facilities lacking sophisticated deterministic lithography infrastructure.

\subsection{Demonstration of scalability across an array of devices}\label{subsec5}
To assess the integration accuracy between the CBG resonators and the embedded SCQDs, we performed low-temperature CL mapping on a $6\times6$ array of SCQD-CBGs. Low-temperature (20~K) CL mapping of the full $6 \times 6$ SCQD-CBG array (Fig. \ref{fig:2} (a)) reveals bright emission strictly confined to the CBG-mesa centers, confirming the precision of SCQD nucleation and their robust integration with the CBGs. Moreover, the emission wavelengths of all devices are spectrally homogeneous within $(934\pm6)$~nm, highlighting not only the reproducibility of the growth-fabrication scheme but also its scalability toward large, narrow-wavelength-range SPS arrays. Noteworthy, the reported spectral inhomogeneity lies in the range accessible by quantum confined Stark tuning~\cite{Patel2010, Bennett2010}, which is compatible with the CBG concept~\cite{Wijitpatima2024}. The corresponding CL spectra for the device array shown in the CL maps are provided in the SI Section 4 (Fig.~S 4).

\subsubsection{Assessment of alignment accuracy}\label{subsubsec1}

\begin{figure}
 \centering
  \includegraphics[width=1.0\textwidth]{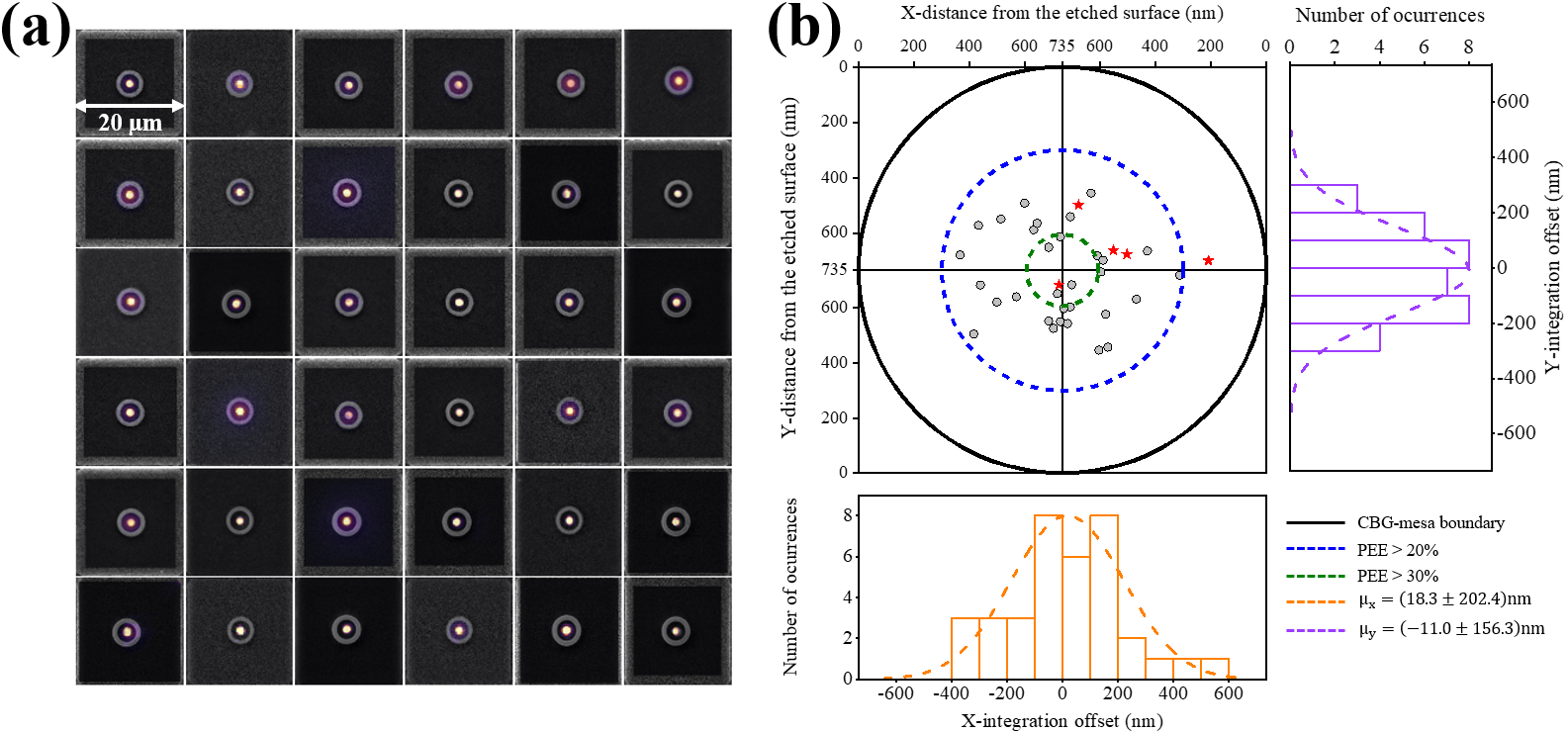}
  \caption{ (a) Low-temperature ($20$~K) CL map (overlaid with SEM image) of a $6\times6$ SCQD-CBG device array (spectral range: $(934\pm6)$~nm), showing emission confined to the CBG-mesa centers with a 100\% integration yield. (b) Statistical analysis of alignment accuracy from high-resolution CL scans (50~nm pixel size) of all 36 devices. QD positions were localized by 2D Gaussian fitting and compared with CBG-mesa centers extracted from SEM data. The resulting radial offset distribution, with mean displacements of $\mu_x = (18.3\pm202.4)$~nm and $\mu_y = (-11.0\pm156.3)$~nm, confirms the high reproducibility of the SCQD-CBG alignment across the array and validates the robustness of our marker-free fabrication strategy.
}
  \label{fig:2}
\end{figure}

To quantify the alignment accuracy, high-resolution CL maps with 50~nm pixel size were analyzed within a statistical evaluation using a two-step method: a 2D Gaussian fit was applied to localize the QD emission within the CL data, while the CBG center was extracted from simultaneously acquired SEM images by identifying the CBG-mesa edges~\cite{barua2025}. The resulting distribution of radial offsets provides a direct measure of the alignment accuracy, as shown in Fig. \ref{fig:2} (b), and yields mean displacements of $\mu_x = (18.3 \pm 202.4)$ nm and $\mu_y = (-11.0 \pm 156.3)$ nm. Applied across all 36 devices, this method showed revealed 100\% process yield in the sense that every SCQD was located within the central 1.47 $\mu$m-diameter CBG-mesa. Notably, 5 devices exhibited a radial offset within $\pm$125 nm, corresponding to a simulated PEE $>$ 30\%, and 30 devices fell within $\pm$435 nm, yielding $>$ 20\% PEE. While deterministic alignment techniques based on QD pre-localization have achieved sub-50~nm accuracies in optimized workflows~\cite{madigawa2025, Rickert2025, barua2025}, the results presented here are particularly significant as the entire integration process was carried out without the use of alignment markers, in-situ patterning, or prior emitter localization by CL or PL imaging. The observed alignment thus stems from a dual precision, intrinsic SCQD nucleation at the CBG-mesa center and photonic cavity placement registered to the SCG-mesas. This level of marker-free integration accuracy highlights a practical and scalable pathway to realizing high-performance single-photon devices with reproducible efficiency, positioning this approach as a compelling alternative to deterministic alignment methods for large-scale quantum photonic architectures.

\subsection{Systematic characterization of SCQD-CBGs}

A complete optical and quantum optical investigation was carried out on a selected subset of five SCQD-CBG devices from a $6\times6$ SCQD-CBG array, with systematically varied radial offsets between the SCQDs and the mesa center: SCQD-CBG1 (527~nm), SCQD-CBG2 (243~nm), SCQD-CBG3 (240~nm), SCQD-CBG4 (197~nm), and SCQD-CBG5 (54~nm) (indicated by red stars in Fig. \ref{fig:2} (b)). All measurements were performed on the charged exciton ($X^{-}$) emission line of each device, since it consistently appeared as the most dominant and intense transition. Comprehensive characterization, including $\mu$PL, time-resolved lifetime measurements, second-order autocorrelation, Hong-Ou-Mandel (HOM) two-photon interference (TPI), as well as high-resolution linewidth determination via Fabry-Pérot interferometry (FPI), was conducted on all five devices, providing a complete assessment of their optical and quantum-optical properties and the impact of emitter radial displacement from the CBG-mesa center. A full description of the optical setup and measurement parameters, along with its schematic diagram, is provided in the SI (Section 1). Details and raw datasets of all five SCQD-CBG devices are provided in the SI (Section 5).

\subsubsection{Optical characterization}\label{subsubsec3}
\begin{figure}
 \centering
  \includegraphics[width=1.0\textwidth]{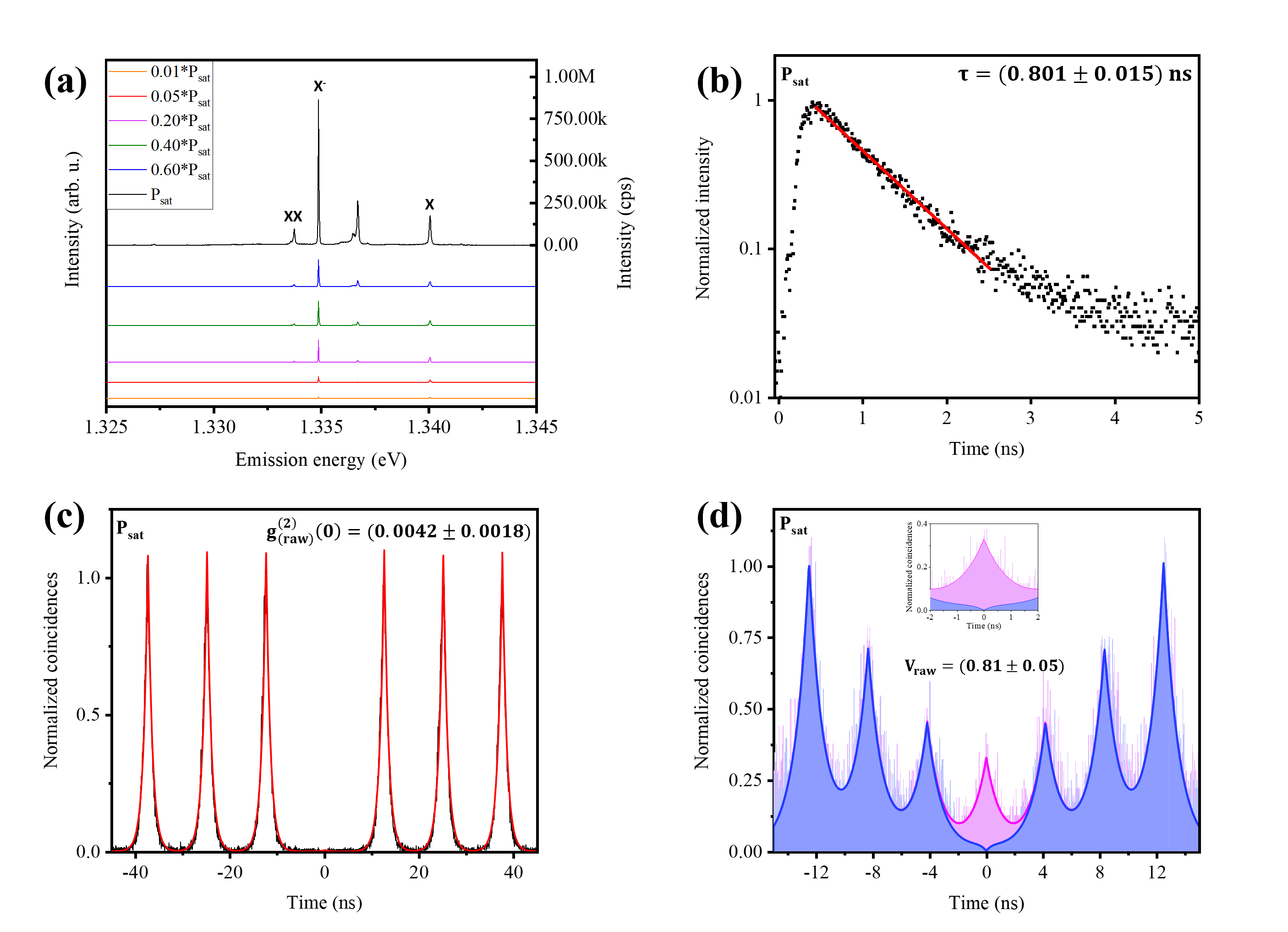}  
  \caption{(a-d) Optical and quantum optical characterization of the best-performing SCQD-CBG5 device (radial offset: 54~nm), measured at 4~K. (a) $\mu$PL spectra under pulsed excitation at 870~nm for varying pump powers (vertically offset for clarity), revealing a max PEE of $(47.1\pm3.8)\%$. All subsequent measurements were performed under p-shell excitation at 911~nm and at the QD saturation power. (b) Time-resolved $\mu$PL indicating a QD radiative decay lifetime of $(0.80\pm0.02)$~ns. (c) Second-order autocorrelation measurements demonstrating $g^{(2)}(0)=(0.0042\pm0.0018)$, corresponding to a single-photon purity of $(99.58\pm0.18)$\%. (d) HOM TPI measurements exhibiting a raw visibility of $(81\pm5)$\%, evidencing very high photon indistinguishability.
}
  \label{fig:4}
\end{figure}

To evaluate the optical performance of the devices as a function of QD position, we conducted $\mu$PL characterization on the five selected SCQD-CBG devices. Quasi-resonant excitation was performed using a tunable pulsed laser set to 870~nm. Corresponding power-dependent $\mu$PL spectra (Fig. \ref{fig:4} (a) were recorded for the best-performing device (radial offset = 54~nm). This device, SCQD-CBG5, delivered the highest brightness with a maximum detected count rate of 2.73~MHz at saturation. After correcting for the independently calibrated setup efficiency of $(7.3\pm0.3)\%$ (see SI, Section 1), this corresponds to a PEE of $(47.1\pm3.8)\%$. To our knowledge, this is the highest value of PEE achieved for an SCQD integrated into a nanophotonic cavity, underscoring the effectiveness of our marker-free approach in combination with the CBG concept. This record value confirms that near-ideal alignment can be achieved without complex deterministic lithography or pick-and-place techniques, establishing a benchmark for scalable device fabrication. The raw $\mu$PL spectra at saturation powers of all five SCQD–CBG devices are provided in the SI, Fig. S 5.

\subsubsection{Time-resolved PL studies}\label{subsubsec4}
Time-resolved $\mu$PL under quasi-resonant $p$-shell excitation was employed to evaluate the radiative lifetimes of the five selected SCQD-CBG devices, yielding values of $(0.80\pm0.02)$~ns, $(0.70\pm0.01)$~ns, $(0.92\pm0.01)$~ns, $(1.01\pm0.02)$~ns, and $(1.02\pm0.02)$~ns for SCQD-CBG5 through SCQD-CBG1, respectively. Fig. \ref{fig:4} (b) displays the radiative decay lifetime of SCQD-CBG5. These numbers fall within the expected range for (buried-stressor) InGaAs QDs~\cite{Wang1994, Yu1996, MaxStrauss2017} and show no systematic dependence on radial offset. As the CBGs were optimized for high PEE rather than Purcell enhancement, the absence of any systematic lifetime variation with radial displacement indicates that the radiative dynamics are governed predominantly by the intrinsic QD properties. This conclusion is corroborated by our FEM analysis (SI Section 3), which yields a near-unity Purcell factor (between 0.9 and 1.3) over the SCQD emission band of 
$(934\pm6)$~nm, demonstrating that the CBG structure introduces no measurable spontaneous-emission enhancement. This also confirms that PEE can be engineered independently of lifetime modification, offering design freedom for future structures that may selectively target brightness or Purcell enhancement. Furthermore, the narrow lifetime spread across the devices suggests strong homogeneity of the buried-stressor QD ensemble, which is advantageous for reproducibility across large arrays. Time-resolved decay traces of the five SCQD-CBG devices are shown in the SI, Fig. S 6.

\subsubsection{Single-photon emission purity}\label{subsubsec5}
The single-photon nature of emission from the SCQD-CBG devices was rigorously assessed through second-order autocorrelation measurements using an HBT setup. All measurements were performed at the respective saturation powers of the QDs, where multi-photon generation and background contributions are typically most pronounced. Excitation into the p-shell of the QD was employed to reduce carrier relaxation time jitter while avoiding laser stray light, thus allowing for clean isolation of the QD s-shell emission. The filtered emission was directed to a pair of SNSPDs, and coincidence histograms were acquired to extract the zero-delay second-order correlation \(g^{(2)}(0)\), serving as a direct measure of the single-photon purity. Across the five devices (SCQD-CBG5-1), we observe consistently strong photon antibunching with \(g^{(2)}(0)\) values of $(0.4\pm0.2)$\%, $(0.6\pm0.3)$\%, $(2.6\pm0.7)$\%, $(2.3\pm0.7)$\%, and $(2.7\pm0.3)$\%. Notably, the most spatially centered emitter exhibits a \(g^{(2)}(0)=(0.0042\pm0.0018)\), corresponding to a single-photon purity of $(99.58\pm0.18)$\%, highlighting the intrinsic suppression of multiphoton events even at QD saturation power (Fig. \ref{fig:4} (c)). This value represents, to the best of our knowledge, the lowest multiphoton probability reported for any epitaxial SCQD system. Importantly, the measured 
\(g^{(2)}(0)\) values exhibit no discernible dependence on emitter radial displacement from the CBG center, confirming that the suppression of multi-photon events is governed by intrinsic QD recombination dynamics rather than spatial coupling efficiency. The consistently sub-3\% values across all structures affirm the robustness of the SCQD growth platform and the clean photonic environment of the CBG, both of which minimize spurious emission channels such as background luminescence or emission from adjacent states. These results underline the scalability of our integration scheme, demonstrating high-purity single-photon generation without reliance on resonant pumping or spectral filtering, and establish the platform as a viable alternative to more complex deterministic or in-situ aligned systems. SI Fig. S 7 presents the raw \(g^{(2)}(\tau)\) histograms for all five SCQD-CBG devices.

\subsubsection{Photon extraction efficiency}\label{subsubsec6}

\begin{figure}
 \centering
  \includegraphics[width=1.0\textwidth]{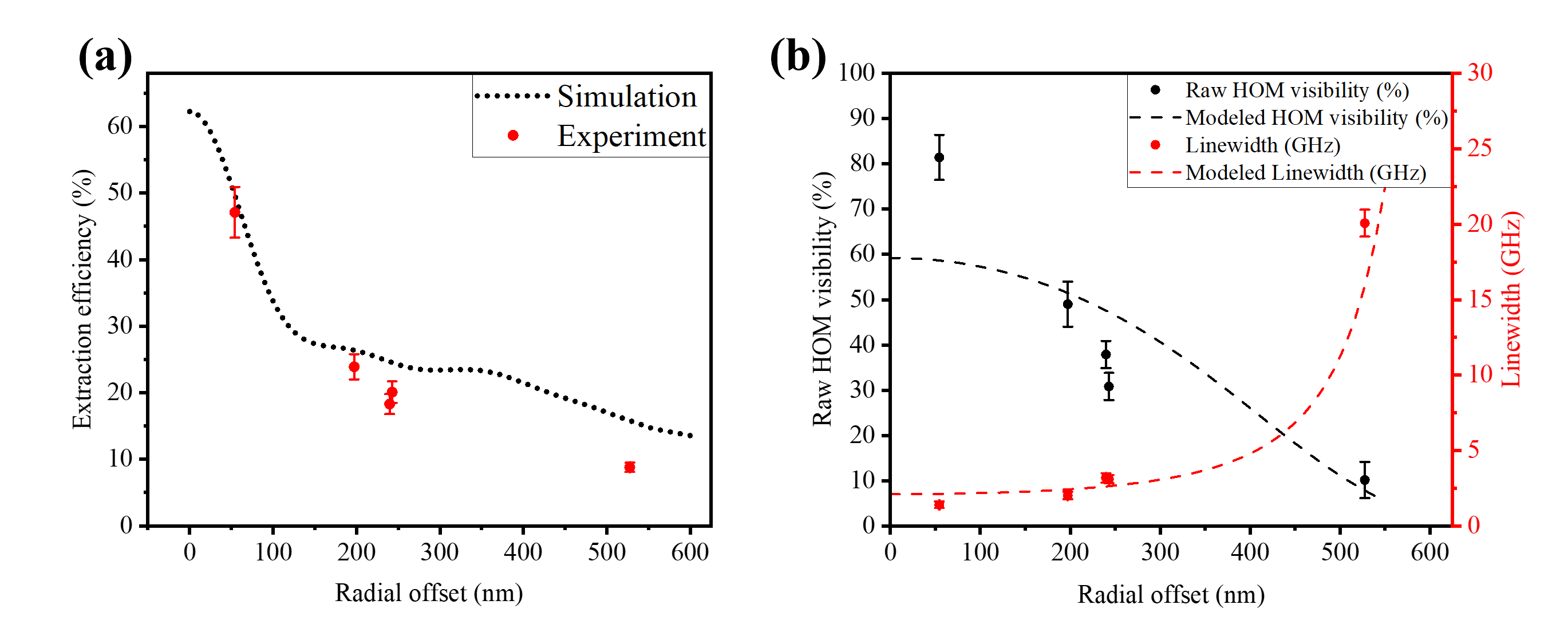}
  \caption{Assessment of optical and quantum optical device performance as a function of radial offset. (a) QD radial-offset dependence of PEE, with FEM simulations (black dashed line) in agreement with experimental results (red circles). (b) Dependence of raw HOM-TPI visibility (\%, left axis) and emission linewidth (GHz, right axis) on the QD radial offset. Experimental data are represented by black (visibility) and red (linewidth) circles, while the corresponding theoretical modeling is depicted by black and red dashed lines.
}
  \label{fig:5}
\end{figure}

To rigorously quantify the role of spatial alignment, we systematically correlated experimentally extracted PEEs from SCQD-CBG devices with varying radial offsets (red circles in Fig. \ref{fig:5} (a)) to FEM simulations (black dashed line). The optimally aligned device, SCQD-CBG5 (54~nm radial offset) already delivers a PEE of $(47.1\pm3.8)\%$, in close agreement with FEM simulations predicting 50.2\% at this displacement and a theoretical maximum of 62.2\% for a perfectly centered emitter. Devices with larger radial offsets (SCQD-CBG4-1) exhibit reduced [simulated] efficiencies of $(23.9\pm1.9)\%$ [26.4\%], $(18.3\pm1.5)$\% [24.7\%], $(20.1\pm1.6)$\% [24.6\%], and $(8.8\pm0.7)$\% [15.9\%], reflecting the expected decline in emitter-resonator coupling due to misalignment. The overall experimental trend matches the simulated results, with minor deviations attributable to fabrication tolerances and local non-uniformities. This agreement between experiment and simulation establishes a quantitative benchmark for alignment tolerances in SCQD-CBG devices.

\subsubsection{Linewidth characterization using Fabry-Pérot interferometry}\label{subsubsec7}

Spectral linewidth measurements offer crucial insight into pure dephasing processes and play a central role in determining the suitability of QDs for SPS-based protocols. Under quasi-resonant p-shell excitation at saturation power, we employed a scanning FPI with a spectral resolution of 150~MHz to resolve the emission linewidths of the SCQD-CBG devices. Noteworthy, four out of five devices exhibited spectrometer resolution-limited ($\sim$30~$\mu$eV) linewidths, necessitating high-resolution FPI analysis. The most optimally aligned structure (SCQD-CBG5) demonstrated an emission linewidth of $(1.41\pm0.22)$~GHz, a remarkable result considering the complete absence of electrical or charge-noise suppression mechanisms, which are often used to reduce spectral jitter~\cite{Kuhlmann2013, Somaschi2016}. The other four devices yielded linewidths of $(2.03\pm0.25)$\%, $(3.20\pm0.31)$, and $(3.12\pm0.27)$~GHz, respectively. In contrast, the device with the largest radial displacement (527~nm), located $\sim$200~nm away from the etched boundary of the CBG mesa, exhibited significant broadening to $(20.07\pm0.89)$~GHz, consistent with increased sensitivity to surface charge noise~\cite{Connors2019}. These findings establish a clear correlation between emitter-resonator alignment and spectral purity, reinforcing the importance of spatial precision in minimizing environmental noise and optimizing optical quality in integrating SPS. 
\\
Several findings \cite{Connors2019, Manna2020} suggest that charge fluctuations caused by surface states at the etched surface of the CBG mesa play a major role in causing both decoherence and linewidth broadening. These fluctuating charges generate local electric fields that cause a time-varying energy shift in the emission frequency, leading to linewidth broadening. Since these fields decay with inverse distance squared, QDs with a higher radial displacement are affected more strongly by these fields, causing the aforementioned dependence of the linewidth on the displacement. As derived in the SI, Section 6, the linewidth broadening scales like
\begin{align}
  \gamma_\mathrm{deph}(R) = c \, \frac{R_0^2 + R^2}{(R_0^2 - R^2)^3},
\end{align}
where $R$ is the radial displacement, $R_0$ is the radius of the CBG mesa, and $c$ is a proportionality factor depending on the charge fluctuator density on the mesa interface. We stress that this relationship should be understood as describing the average behavior over many emitters, as it assumes a uniform distribution of surface charge states; in practice, the charge distribution can vary from sample to sample. Nevertheless, the theory curve in Fig.~\ref{fig:5} (b) agrees very well with the measured linewidths of the five samples, using $c=\SI{0.220}{\nano \second^{-1} \micro \meter^{4}}$ as the only fitting parameter.

\subsubsection{Photon indistinguishability}\label{subsubsec8}

To directly assess the influence of spatial emitter placement on photon indistinguishability, we conducted HOM TPI measurements under pulsed quasi-resonant p-shell excitation, using a fixed interferometric delay of 4~ns and operating each SCQD at its saturation power. The results are striking that for the most precisely centered emitter (radially offset by only 54~nm), a raw HOM visibility of $(81\pm5)$\% (see Fig. \ref{fig:4} (d)) is achieved, a record-breaking value for epitaxial SCQD systems and comparable to the best demonstrations using self-assembled QDs under p-shell excitation. For increasing lateral displacement we observe a clear degradation in indistinguishability, with visibilities dropping to $(49\pm5)$\%, $(38\pm3)$\%, $(31\pm3)$\%, and as low as $(10\pm4)$\% (the raw coincidence histograms from HOM TPI measurements on all five devices are included in SI Fig. S 8.) in the most extreme case of  SCQD-CBG1 with a lateral displacement of 527~nm. The latter corresponds to the emitter located $\sim$200~nm away from the etched outer boundary of the CBG mesa, where decoherence mechanisms such as charge noise~\cite{Mudi2025} and surface-induced spectral fluctuations likely dominate in accordance with our theoretical description of the offset-dependent emission linewidth. To our knowledge, this is the first direct and quantitative demonstration of HOM TPI visibility as a function of emitter displacement within a photonic resonator, which provides a valuable basis for further device optimization. The ability to reach such high values of indistinguishability without complex tuning or charge suppression and using a fully marker-free and site-controlled growth approach sets a new benchmark for integrated quantum emitters and opens up a clear path toward scalable quantum light sources for quantum technologies. 
\\
To obtain a better understanding of the underlying physics and limitations of the indistinguishability, we established a theoretical description based on a Markovian dephasing model. Within this approach, the Hong-Ou-Mandel indistinguishability is expressed as \cite{Kaer2013, Grange2015}
\begin{align} \label{eq:indist}
  \mathcal{I} = \frac{\Gamma_\mathrm{rad}}{\Gamma_\mathrm{rad} + \gamma_\mathrm{deph}}.
\end{align}
To get a general trend as a function of emitter displacement, we evaluate Eq.~\eqref{eq:indist} for the previously fitted linewidth  $\gamma_\mathrm{deph} (R) + \Gamma_\mathrm{rad}$ depicted in Fig.~\ref{fig:5} (b) and a constant $\Gamma_\mathrm{rad} = 1.25\, \mathrm{ns}^{-1}$. In the same figure, we also show the indistinguishability that we obtain from our model. The theoretical values match well with the measurement results, especially at higher displacements, and capture the overall trend that indistinguishability improves with better center-positioning.  
The slight underestimation at small displacements results from the assumption of a uniform charge distribution that does not account for sample-specific variations in the local charge environment. In that sense, the results should be interpreted as describing an average behavior over many devices, rather than predictions for an individual sample. In addition, at this stage, we have neglected the effects of dielectric screening, meaning that charge-noise effects in proximity to the center should be overestimated in comparison to off-center positions, which might be another reason for the low HOM visibility predicted by this model.
Regardless of the numerical deviations, the correct reproduction of the observed trends by the model calculations suggests the validity of the underlying assumption that charge noise mainly originates from the etched surface and can, consequently, be optimized by accurate emitter positioning. Notably, the calculations further reveal a saturation of the indistinguishability for $R \leq 50 \, \mathrm{nm}$, indicating that further improvements in spatial alignment are unlikely to significantly improve optical performance.

\section{Discussion and Conclusion}\label{sec13}

The results presented in this work establish a scalable and lithography-compatible route for integrating SCQDs into CBG resonators through a fully marker-free process. Built on a buried-stressor epitaxial platform and automated EBL patterning, the approach achieves sub-500 nm emitter-resonator alignment across an extended array without relying on in-situ lithography or marker alignment. Correlative structural, spectroscopic, and quantum optical analyses reveal a clear correspondence between spatial offset and device performance, providing a unified experimental-numerical framework to quantify spatial-spectral coupling in quantum emitters.

\begin{table}[h!]
\centering
\footnotesize
\caption{Summary of performance metrics for five SCQD-CBG devices with varying radial offsets.}
Abbreviations: $\Delta R$: Radial offset, PEE: Photon extraction efficiency, L: Lifetime, LW: Linewidth, V: Photon indistinguishability.
\label{tab:device_performance}
\begin{tabular}{|c|c|c|c|c|c|c|}
\hline
\textbf{Structure} & \textbf{$\Delta R$ (nm)} & \textbf{PEE (\%)} & \textbf{L (ns)} & \textbf{$g^{(2)}(0)$ (\%)}& \textbf{LW (GHz)} & \textbf{V (\%)} \\
\hline
SCQD–CBG1 & 527  &  $(8.8\pm0.7)$  & $(1.02\pm0.02)$ & $(2.7\pm0.3)$ & $(20.07\pm0.89)$ & $(10\pm4)$  \\
SCQD–CBG2 & 243  & $(20.1\pm1.6)$  & $(1.01\pm0.02)$ & $(2.3\pm0.7)$ & $(3.12\pm0.27)$ & $(31\pm3)$  \\
SCQD–CBG3 & 240  & $(18.3\pm1.5)$  & $(0.92\pm0.01)$ & $(2.6\pm0.7)$ & $(3.20\pm0.31)$ & $(38\pm3)$  \\
SCQD–CBG4 & 197  & $(23.9\pm1.9)$  & $(0.70\pm0.01)$ & $(0.6\pm0.3)$ & $(2.03\pm0.25)$ & $(49\pm5)$  \\
SCQD–CBG5 & 54   & $(47.1\pm3.8)$  & $(0.80\pm0.02)$ & $(0.4\pm0.2)$ & $(1.41\pm0.22)$ & $(81\pm5)$  \\
\hline
\end{tabular}
\end{table}

Table~\ref{tab:device_performance} summarizes the performance metrics of five representative SCQD–CBG devices with increasing radial offset. The data confirm that emitter–resonator displacement primarily governs the optical coupling strength, directly impacting PEE, linewidth, and HOM TPI visibility, while intrinsic properties such as lifetime and single-photon purity remain largely unaffected. The best-performing device exhibits a PEE of $(47.1\pm3.8)\%$, single-photon purity of $(99.58\pm0.18)\%$, linewidth of $(1.41\pm0.22)$~GHz, and HOM visibility of $(81\pm5)\%$, setting a new benchmark for SCQD-based sources. These results demonstrate that epitaxial precision, combined with optimized resonator design, can yield high brightness and photon indistinguishability, both of which are essential for scalable quantum photonic architectures.

\begin{table}[h!]
\centering
\footnotesize
\caption{Overview of state-of-the-art QD alignment approaches and device performance metrics.}
Abbreviations: QD: Quantum dot growth method, $\Delta R$: Radial offset, IT: Integration technique, WL: Wavelength, PEE: Photon extraction efficiency, LW: Linewidth, V: Photon indistinguishability, Ref: Reference, DE: Droplet etched quantum dots, SK: Stranski-Krastanov grown quantum dots, SCQD: Site-controlled quantum dots, NH: Nanoholes, NW: Nanowires, BS: Buried-stressor quantum dot growth, CBG: Circular Bragg grating resonators, MB: Marker-based integration, iEBL: In-situ electron beam lithography, C: Conventional EBL integration (not using alignment markers or in-situ techniques).
\label{tab:state of the art}
\begin{tabular}{|c|c|c|c|c|c|c|c|c|c|}
\hline
\textbf{QD} & \textbf{Structure} & \textbf{$\Delta R$ (nm)} & \textbf{IT} & \textbf{WL (nm)} & \textbf{PEE (\%)} & \textbf{LW ($\mu$eV)} & \textbf{$g^{(2)}(0)$}& \textbf{V (\%)} & \textbf{Ref} \\
\hline
DE & Pillar+rings  &  50  & MB & 780 & 39 &- & 0.024 & 21 & \cite{madigawa2025} \\
DE & Hybrid-CBG  & 66  & iEBL & 895 & 68 &30& 0.011 & 53 & \cite{barua2025}\\
SK & Hybrid-CBG  & 32  & MB & 940 & - &8& 0.086 & 96 & \cite{Rickert2025}\\
NH-SCQD &-& 80  &  -  & 910 & - & 7 & 0.020 & 73 & \cite{Jons2013}  \\
NW-SCQD &-& 200  &- & 880  & 42 & - & 0.120 & - & \cite{Reimer2012}  \\
BS-SCQD &Microlens& -  & iEBL & 950  & 21 & 30 & 0.030 & - & \cite{ArsentyElsevier}  \\
BS-SCQD &-& -  & - & 930  & - & 27 & 0.026 & 65 & \cite{JanAPLPhotonics2020}  \\
BS-SCQD &CBG& 54  & C & 930  & 47 & 6 & 0.004 & 81 & This work  \\
\hline
\end{tabular}
\item[*] Quantum optical measurements in Refs.~\cite{Reimer2012, JanAPLPhotonics2020} were performed under continuous-wave (CW) excitation conditions.
\end{table}

A broader comparison with state-of-the-art alignment approaches is provided in Table~\ref{tab:state of the art}. Despite eliminating alignment markers and an in-situ approach, the present platform achieves optical and quantum optical performance comparable to, and in several aspects exceeding, the results obtained using such complex schemes. In particular, the combination of high PEE, narrow linewidth, and high HOM visibility at QD saturation power under quasi-resonant excitation parallels that of marker-based SK and droplet-etched systems, while maintaining full compatibility with planar wafer-scale processing. The reproducibility and stability observed across multiple devices further underline the precision of the buried-stressor growth mode and the robustness of automated CBG patterning.

Together, these results establish a fabrication-efficient pathway toward scalable arrays of SPS with high indistinguishability. Looking ahead, the platform naturally extends toward wafer-scale arrays of SCQD-CBG devices, where Stark shift~\cite{Patel2010} or strain engineering~\cite{Rota2024} can spectrally tune distinct emitters to a common target resonance. Such fully tunable arrays open pathways to on-chip SPS with high indistinguishability for quantum networks~\cite{Zhai2022, Pont2025}, as well as application-specific architectures such as neuromorphic photonic processors~\cite{Farmakidis2024, Li2025}. The demonstrated alignment tolerance and fabrication robustness lay the groundwork for integrating large numbers of sources into complex photonic circuits, ultimately bridging the gap between proof-of-concept quantum light sources and large-scale, application-ready quantum photonic hardware.

\backmatter

\bmhead{Supplementary information}

The Supplementary Information provides comprehensive details of the experimental setup, optical characterization of SCG-mesas, design parameters of the CBG resonators, CL spectra of the complete device array, raw optical and quantum optical data from CBGs with varying radial offsets, and the modeling framework used to evaluate charge noise amplitude. 

\bmhead{Acknowledgements}

The authors thank Ching-Wen Shih, Lukas Dworaczek, Martin Podhorský, and Maximilian Klonz for their invaluable input in scientific discussions. The authors thank Lucas Rickert and Setthanat Wijitpatima for the helpful discussions regarding device fabrication. The authors thank Martin von Helversen for supporting Fabry-Pérot interferometry measurements. The authors would also like to thank Kathrin Schatke, Praphat Sonka, Heike Oppermann, and Stefan Bock for their expert technical support.

\section*{Declarations}

\begin{itemize}
\item Funding:
This work was funded by the German Research Foundation (RE2974/33-1 and INST 131/795-1 320 FUGG), and by the Federal Ministry of Research, Technology and Space (BMFTR) through the projects MultiCoreSPS (Grant No. 16KIS1819K) and QR.N (Grant No. 16KIS2203).

\item Conflict of interest:

The authors declare no conflicts of interest.

\item Data availability:

The data that support the findings of this study are available from the corresponding authors upon reasonable request.
\item Author contribution:

SR conceived the project with critical input from KG and IL. KG carried out sample growth and device fabrication, led methodology development, experimental investigation, and data acquisition and analysis. ST supported device fabrication. AB, ST, AK-S, IL, PM, and SvR supported methodology development, while AB, ST, PM, SB, NN, and CCP contributed to experimental investigations. LJR, with input from SvR, performed the alignment accuracy analysis. SW, together with AS and CG, independently developed the theoretical modeling of offset-dependent linewidth and HOM visibility, with SW also contributing to the theoretical modeling section of the manuscript. SR supervised the project, secured funding, and coordinated project administration. KG and SR led manuscript preparation, with critical contributions from all co-authors.

\end{itemize}

\bibliography{sn-bibliography}

\end{document}


\title[Scalable Quantum Photonic Platform Based on Site-Controlled Quantum Dots Coupled to Circular Bragg Grating Resonators: Supplementary Information]{Scalable Quantum Photonic Platform Based on Site-Controlled Quantum Dots Coupled to Circular Bragg Grating Resonators: Supplementary Information}

\author*[1]{\fnm{Kartik} \sur{Gaur}}\email{kartik.gaur@tu-berlin.de}
\author[1]{\fnm{Avijit} \sur{Barua}}
\author[1]{\fnm{Sarthak} \sur{Tripathi}}
\author[1]{\fnm{Léo J.} \sur{Roche}}
\author[2]{\fnm{Steffen} \sur{Wilksen}}
\author[2]{\fnm{Alexander} \sur{Steinhoff}}
\author[1]{\fnm{Sam} \sur{Baraz}}
\author[1]{\fnm{Neha} \sur{Nitin}}
\author[1]{\fnm{Chirag C.} \sur{Palekar}}
\author[1]{\fnm{Aris} \sur{Koulas-Simos}}
\author[1]{\fnm{Imad} \sur{Limame}}
\author[1]{\fnm{Priyabrata} \sur{Mudi}}
\author[1]{\fnm{Sven} \sur{Rodt}}
\author[2]{\fnm{Christopher} \sur{Gies}}
\author*[1]{\fnm{Stephan} \sur{Reitzenstein}}\email{stephan.reitzenstein@physik.tu-berlin.de}

\affil[1]{\orgdiv{Institut für Physik und Astronomie}, \orgname{Technische Universität Berlin}, \orgaddress{\street{ Hardenbergstraße 36}, \city{Berlin}, \postcode{10623}, \country{Germany}}}

\affil[2]{\orgdiv{Institute for Physics}, \orgname{Carl von Ossietzky Universität Oldenburg}, \orgaddress{\city{Oldenburg}, \postcode{26129}, \country{Germany}}}

\maketitle

\section{Experimental setup}\label{Sup_sec1}
\begin{figure}[hbt!]
 \centering
  \includegraphics[width=1.0\textwidth]{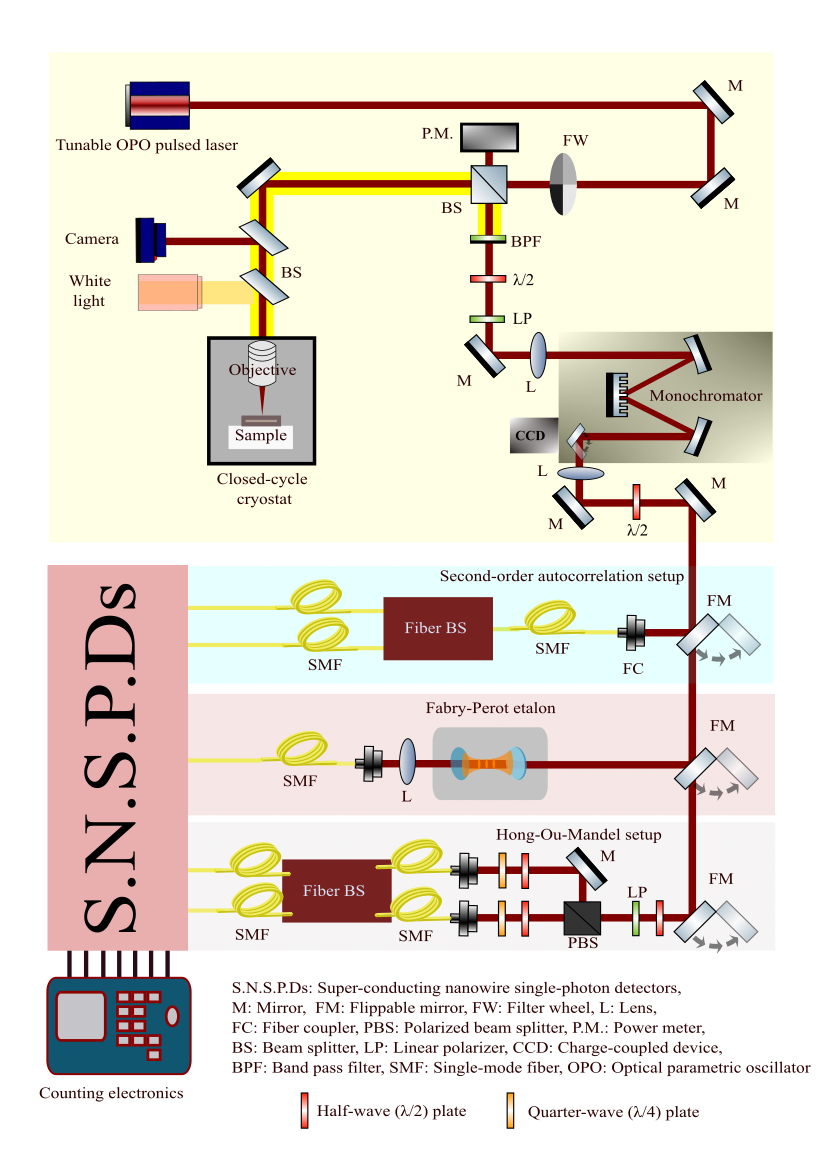}
  \caption{ Schematic representation of the experimental setup.}
 \label{fig:S1}
\end{figure}

All optical experiments were carried out at cryogenic temperatures (T = 4 K) in a closed-cycle cryostat equipped with high-stability nanopositioning stages. For broad pre-characterization of unpatterned SCG mesas, a CW laser diode at 660~nm was employed to excite the SCQDs at the mesa centers across the $5\times5$ mm$^2$ sample piece. A picosecond-pulsed optical parametric oscillator (OPO), operated at a repetition rate of 80 MHz, was tuned to the p-shell resonance of the QDs to ensure quasi-resonant excitation while minimizing excitation-induced dephasing and background fluorescence. The emitted PL was collected through a cold aspheric objective with an NA of 0.81, ensuring efficient collection of the upward-directed emission from the CBG structures. Stray laser light was effectively suppressed using a custom-tilted 950~nm long-pass filter, optimized for QD emission below 950 nm. The spectrally filtered signal was directed to a high-resolution grating spectrometer (1500 lines/mm), yielding a spectral resolution of $\sim$30 $\mu$eV, and subsequently imaged onto a nitrogen-cooled CCD camera for $\mu$PL measurements. For power-dependent and polarization-resolved spectroscopy, a computer-controlled variable neutral density filter wheel and a motorized half-wave plate combined with a polarizer enabled fine-tuned optical attenuation and polarization analysis.

For quantum optical and time-resolved measurements, the filtered emission was coupled into a polarization-maintaining single-mode fiber and routed to various fiber-based interferometric setups. Second-order correlation measurements were performed via a fiber-integrated Hanbury Brown and Twiss (HBT) configuration, comprising a 50:50 beam splitter and two superconducting nanowire single-photon detectors (SNSPDs) with a combined instrument response function (IRF) of ~55 ps. Coincidence detection and correlation histograms were recorded using time-correlated single-photon counting (TCSPC) electronics. For two-photon interference measurements, Hong–Ou–Mandel (HOM) experiments were conducted in a fiber-based polarization interferometer. The OPO repetition rate was down-sampled to 4~ns to isolate individual photon wave packets. A calibrated polarization control stage consisting of quarter- and half-wave plates directed the signal into two optical delay arms, one of which included a 4~ns fiber delay. The two paths were recombined at a 50:50 fiber beam splitter and sent to two SNSPD channels for time-resolved coincidence analysis. High-resolution spectral linewidth measurements were performed using a scanning Fabry–Pérot interferometer (FPI) integrated into the detection path. The FPI, composed of a thermally and mechanically stabilized tunable etalon, featured a free spectral range (FSR) of 12.3 GHz and a resolution of $\sim$150 MHz. The interferometer was scanned at 5 mHz with a modulation amplitude of 500 mV$_{p-p}$, and the transmission signal was coupled into a multimode fiber and detected by a single-photon avalanche diode (SPAD) channel. A schematic representation of the full experimental setup is shown in Fig. S \ref{fig:S1}

The setup efficiency was determined by calibrating the overall detection efficiency of the $\mu$PL system using a Toptica DL 100 CW laser tuned to the QD emission wavelength ($\sim$ 930~nm). This yielded a system efficiency of $(7.3\pm0.3)$\%, accounting for cumulative losses along the collection path, including transmission through optical elements and the CCD's quantum efficiency at the corresponding wavelength~\cite{barua2025}.

\clearpage

\section{Optical assessment of SCG-mesas}\label{Sup_sec2}
\begin{figure}[hbt!]
 \centering
  \includegraphics[width=0.7\textwidth]{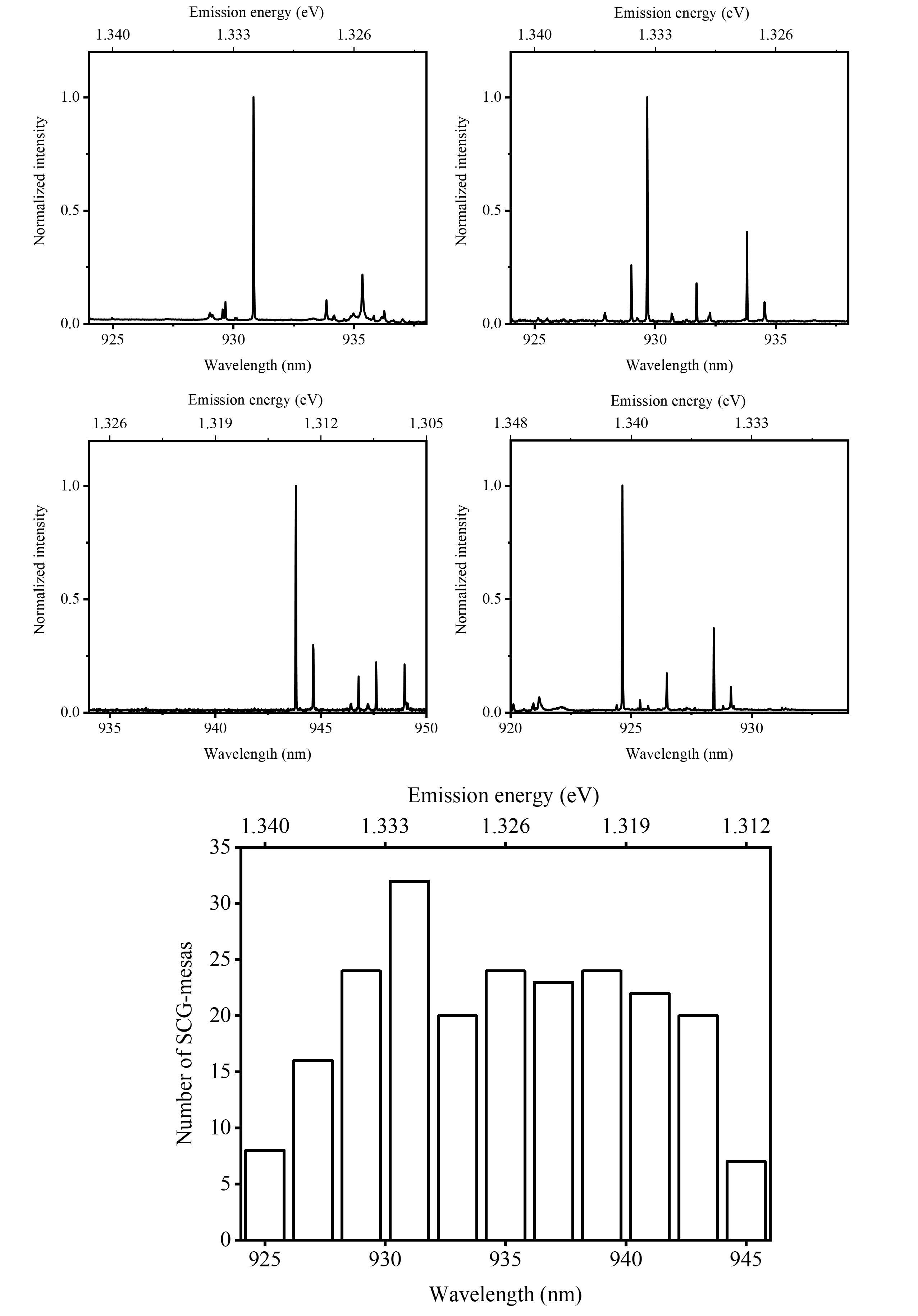}
  \caption{Representative $\mu$PL spectra of SCG-mesas, with the accompanying histogram showing the emission energy distribution across over 200 individual SCG-mesas.}
 \label{fig:S2}
\end{figure}

Optical characterization of the SCG-mesas was performed in above-band excitation $\mu$PL experiments using a CW diode laser at 660~nm. The excitation beam was focused onto the center of each mesa to probe the embedded SCQDs. Emission was collected and analyzed to extract key statistics, including the emission energy distribution, across the $5\times5$ mm$^2$ sample piece. Exemplary emission spectra for individual SCG-mesas are shown (see Fig. S \ref{fig:S2}), accompanied by a histogram summarizing the emission energies of the dominant line from 220 SCG-mesas, providing a quantitative assessment of sample uniformity and spectral reproducibility.

\clearpage

\section{Design and simulation of CBG resonators}\label{Sup_sec3}

\begin{figure}[hbt!]
 \centering
  \includegraphics[width=0.8\textwidth]{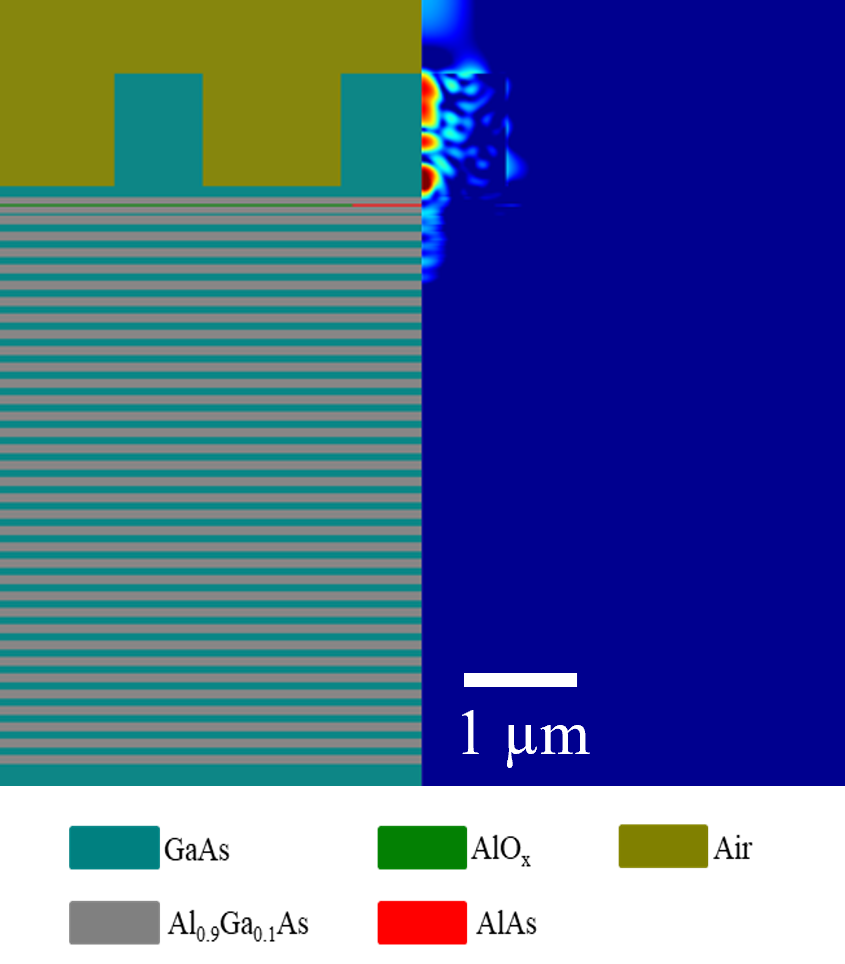}
  \caption{Finite element method (FEM) simulations showcasing the optimized SCQD-CBG geometry (left) and the corresponding electric-field distribution (right).}
  \label{fig:S3}
\end{figure}

\begin{table}[h!]
\centering
\caption{Optimized structural parameters of the fabricated SCQD-CBG devices as used in FEM simulations.}
\label{tab:CBG_parameters}
\begin{tabular}{|c|c|}
\hline
\textbf{Parameter} & \textbf{Value (nm)} \\
\hline
Mesa diameter & 1470 \\
Ring thickness & 740 \\
Ring gap & 1150 \\
Etching depth & 980 \\
Oxidation aperture & 800 \\
\hline
\end{tabular}
\end{table}

\begin{figure}[hbt!]
 \centering
  \includegraphics[width=1.0\textwidth]{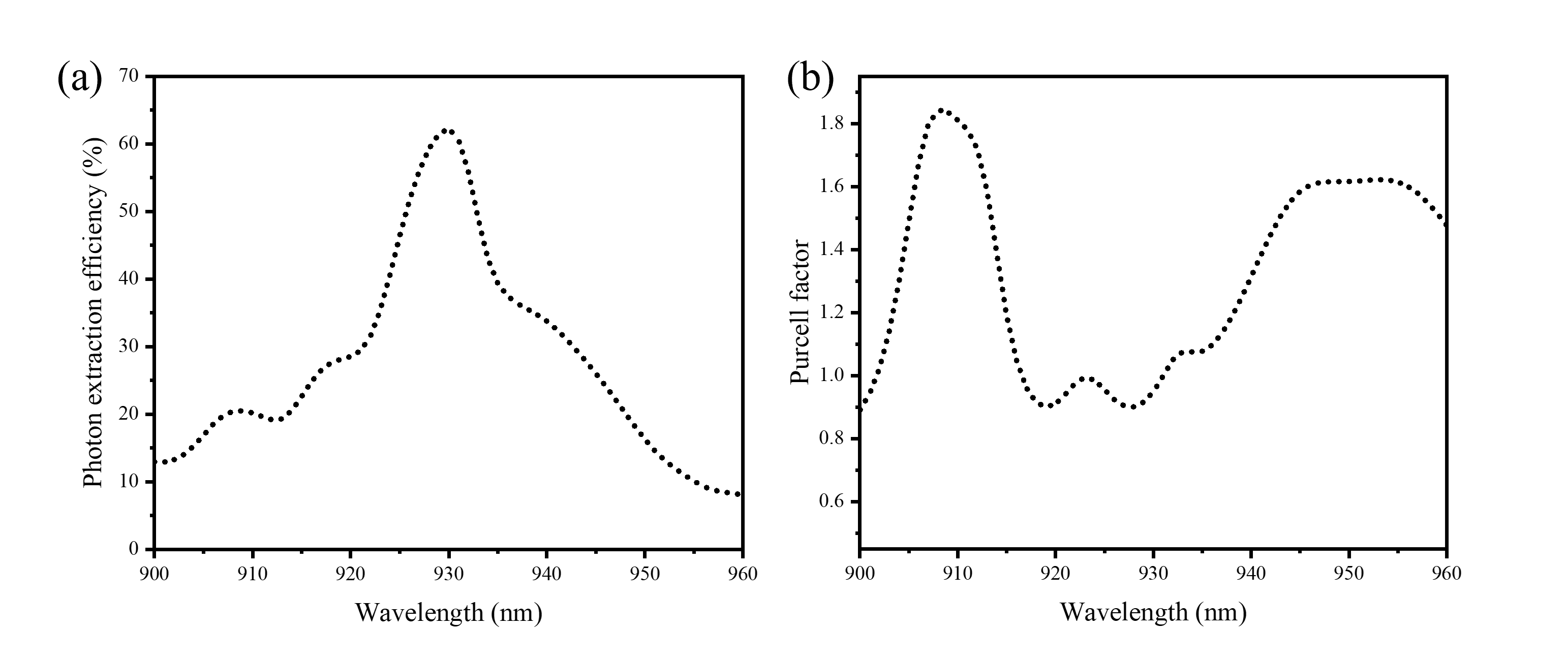}
  \caption{FEM simulation results of the (a) PEE and (b) Purcell factor of the optimized CBG design as a function of dipole emission wavelength.}
  \label{fig:S4}
\end{figure}

In Fig. S \ref{fig:S3}, the left panel shows the simulated cross-section of the optimized SCQD-CBG structure, while the right panel displays the corresponding FEM-calculated electric-field distribution~\cite{JCMsuite}. The simulated electric field distribution shows that the emission originates at the SCQD location and is efficiently redirected upward. The DBR beneath the cavity suppresses downward leakage, leading to a vertically collimated radiation pattern consistent with the high PEE predicted for the optimized design. The final set of optimized structural parameters from FEM simulations, including mesa diameter, etch depth, ring thickness, ring gap, and oxidation aperture, is summarized in Table S \ref{tab:CBG_parameters}. For these design parameters, a spectral sweep of the dipole emission wavelength (from 900 to 960~nm) was performed, revealing a maximum PEE of 62.2\% at the target resonance of 930~nm (Fig. S \ref{fig:S4} (a)), which defines the operational bandwidth and informs tolerance to QD spectral inhomogeneity~\cite{Shih2023}. In the same simulation framework, we also evaluated the expected Purcell enhancement (Fig. S \ref{fig:S4} (b)). Because the structure was optimized solely for maximizing PEE, the simulated Purcell factor remains very close to unity across the relevant spectral window. In particular, within the emission range of our SCQD-CBGs ($(934\pm6)$~nm), the Purcell factor varies only between 0.9 and 1.3. This shallow variation indicates that the CBG introduces no significant modification of the radiative rate, and therefore, no measurable lifetime shortening is expected in the experiments~\cite{Rickert2025}. The experimentally observed lifetimes can thus be interpreted as intrinsic emitter properties, unaffected by cavity-induced enhancement.

\section{Cathodoluminescence emission spectra of the SCQD-CBGs}\label{Sup_sec4}
\begin{figure}[hbt!]
 \centering
  \includegraphics[width=1.0\textwidth]{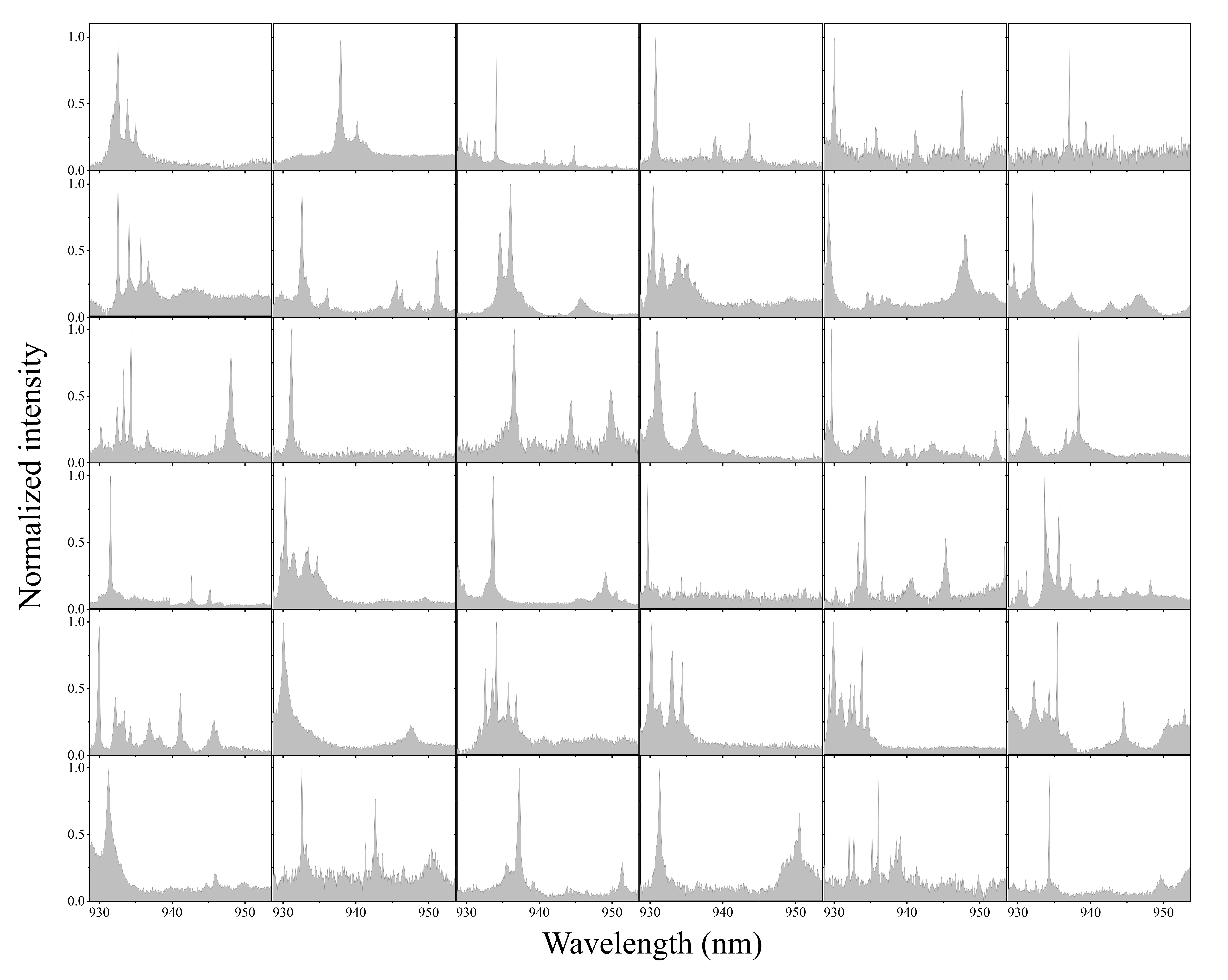}
  \caption{Low-temperature CL spectra corresponding to the $6\times6$ array of SCQD-CBG devices shown in the main text. Each panel displays the emission spectrum acquired from the center of the respective CBG, complementing the CL intensity maps.}
 \label{fig:S5}
\end{figure}

\begin{figure}
[hbt!]
 \centering
  \includegraphics[width=1.0\textwidth]{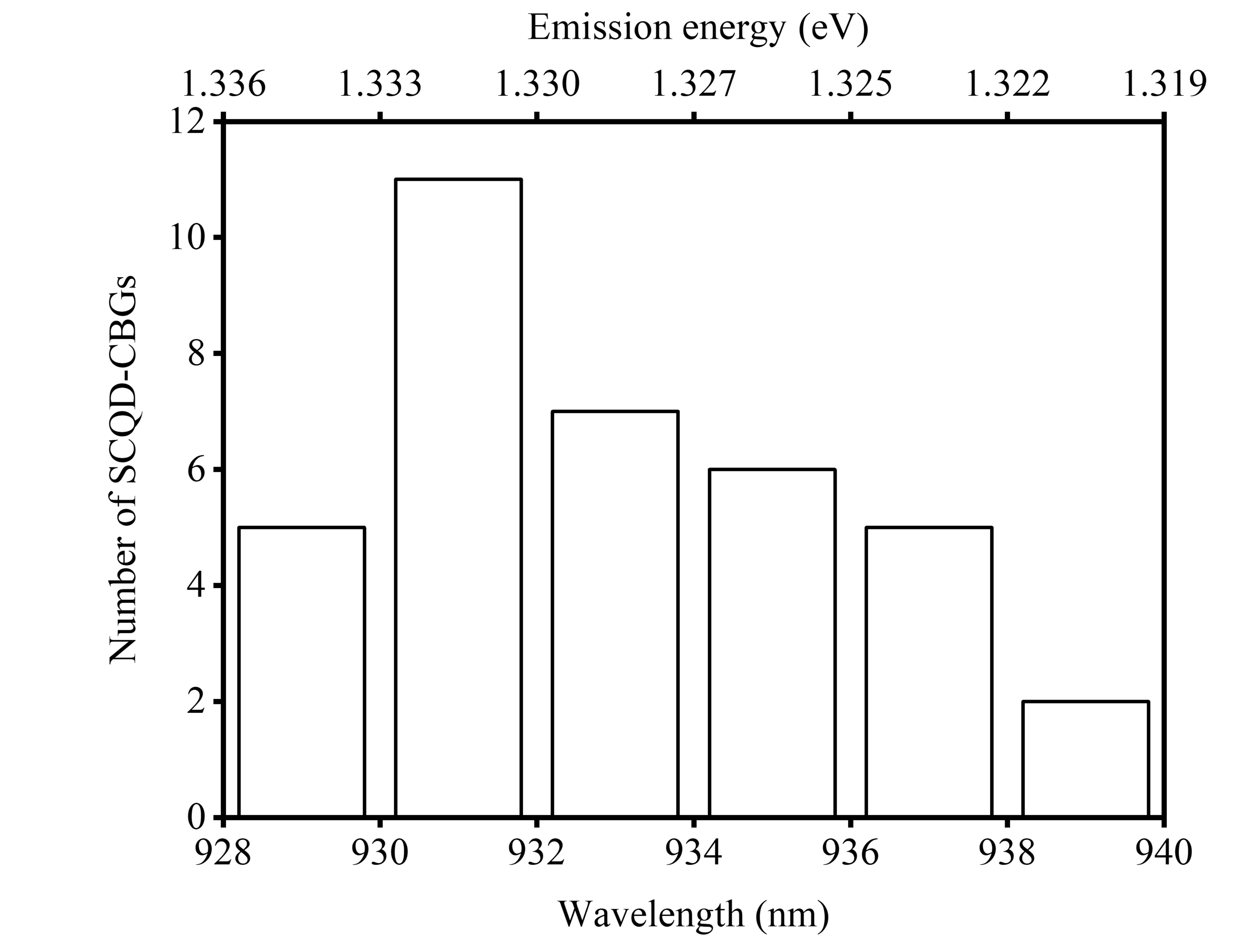}
  \caption{Statistical distribution of the dominant QD emission wavelength extracted from the CL spectra across the $6\times6$ device array.}
 \label{fig:S6}
\end{figure}
As discussed in the main text, low-temperature CL mapping was performed across the complete $6\times6$ array of SCQD-CBG devices to assess spatial alignment and spectral reproducibility. Each device exhibits a localized emission spot at the mesa center, confirming precise positioning of the CBG-mesa with respect to the SCQDs. The corresponding CL spectra from the CBG-mesa centers are shown in Fig. S \ref{fig:S5}. Fig. S \ref{fig:S6} shows the histogram of the dominant emission wavelength obtained from the CL spectra shown above, revealing a narrow spectral spread that reflects the high uniformity of the SCQD emission across the array.

\clearpage

\section{Comparison of devices with varying radial offsets}\label{Sup_sec5}

To assess the influence of the QD-resonator alignment, we investigated five CBGs with different spatial offsets between the emitter position and the mesa center: SC-CBG1 (527 nm), SC-CBG2 (243 nm), SC-CBG3 (240 nm), SC-CBG4 (197 nm), and SC-CBG5 (54 nm), as discussed in the main text. For each device (SC-CBG 1-5), we provide the raw experimental data corresponding to the emission spectra, decay lifetimes, second-order autocorrelation, and HOM TPI. These datasets complement the summarized values presented in the main text.

\subsection{Emission Spectra}\label{Sup_subsec1}
\begin{figure}[hbt!]
 \centering
  \includegraphics[width=0.8\textwidth]{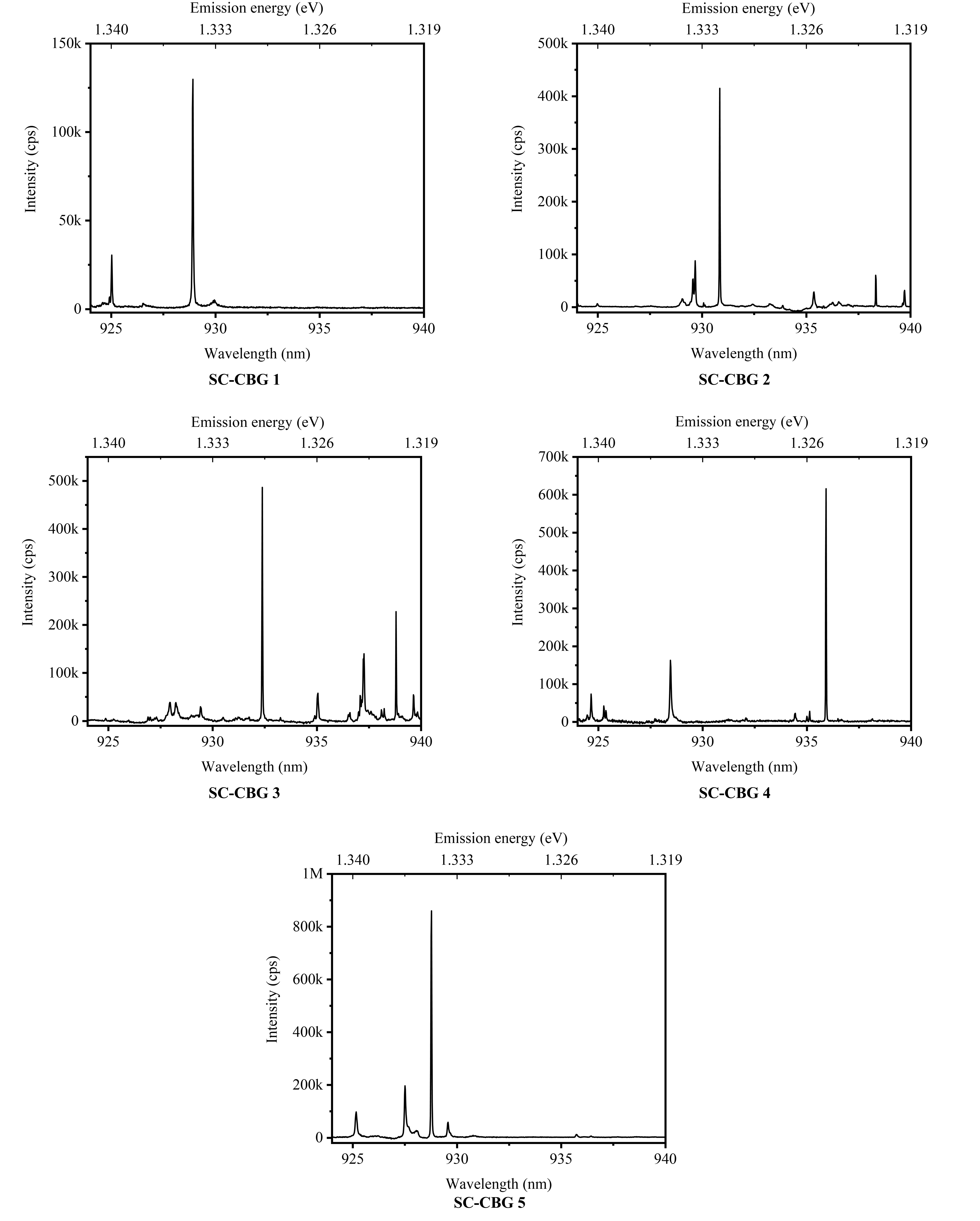}
  \caption{ $\mu$PL spectra of the five devices (SC-CBG 1-5) with different radial offsets.
}
  \label{fig:S7}
\end{figure}

The $\mu$PL spectra of the five devices are displayed in Fig. S \ref{fig:S7}. The spectra confirm single QD emission, with variation in intensity reflecting differences in PEE due to the offset.
\subsection{Lifetimes}\label{Sup_subsec2}
\begin{figure}[h!]
 \centering
  \includegraphics[width=1.0\textwidth]{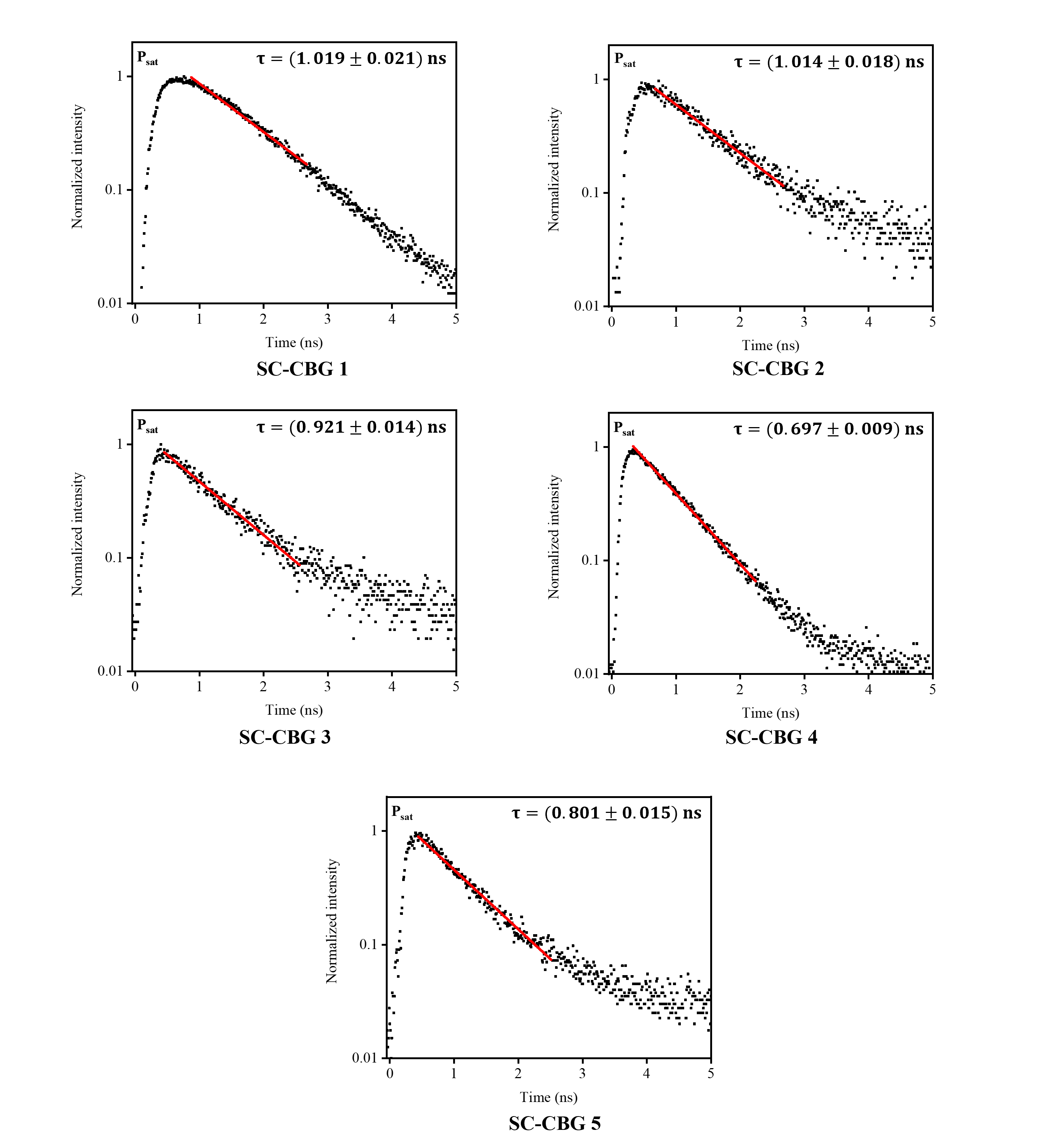}
  \caption{Time-resolved $\mu$PL decay curves of the five devices (SC-CBG 1-5) with different radial offsets.
}
  \label{fig:S8}
\end{figure}

Fig. S \ref{fig:S8} shows the time-resolved $\mu$PL plots for the five SCQD-CBG devices. Exponential fits to the decay yield radiative lifetimes in the range 0.697~ns to 1.019~ns. Importantly, the lifetimes exhibit no systematic dependence on radial offset; the device-to-device spread is comparable to the intrinsic variability of (buried-stressor) InGaAs QDs. The absence of a clear offset-dependent trend indicates a negligible modification of the radiative decay by the CBG structures, consistent with a device design that prioritizes PEE over Purcell enhancement.

\subsection{Second-order autocorrelation}\label{Sup_subsec3}
\begin{figure}[hbt!]
 \centering

  \includegraphics[width=0.93\textwidth]{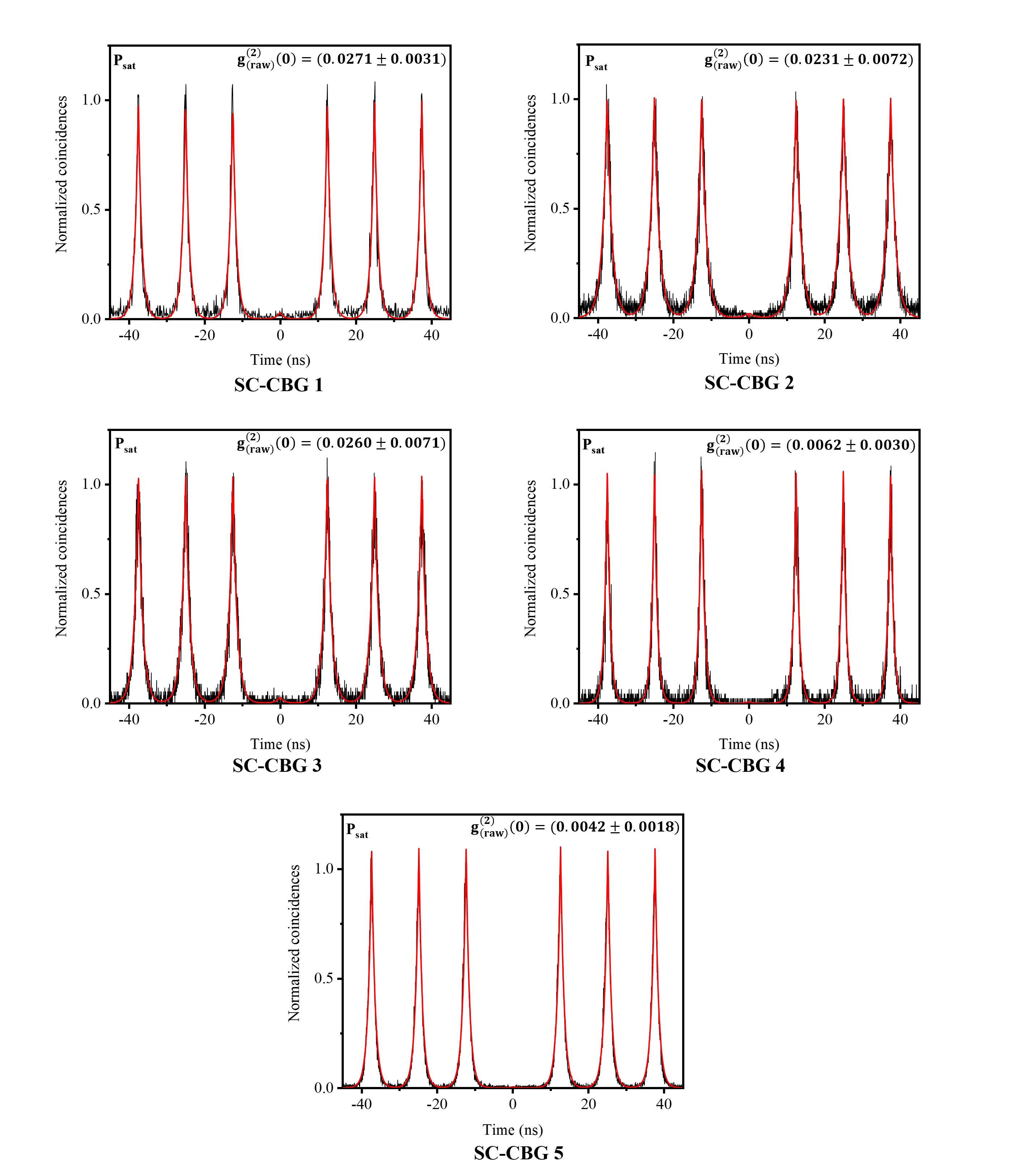}
  \caption{Second-order autocorrelation $g^{(2)}(\tau)$ histograms of the five devices (SC-CBG 1-5) with different radial offsets, all exhibiting pronounced antibunching at zero delay. 
}\label{fig:S9}
\end{figure}
The raw (and fitted) second-order autocorrelation histograms, \(g^{(2)}(\tau)\), for the five devices are shown in Fig. S \ref{fig:S9}. All devices exhibit a pronounced antibunching dip at zero delay, confirming pure single-photon emission. The extracted \(g^{(2)}(0)\) values span a narrow range across the different offsets, with variations attributable to intrinsic emitter properties. Importantly, no systematic dependence on radial displacement is observed, consistent with the expectation that the single-photon purity is governed primarily by the QD itself rather than the CBG alignment.
\clearpage

\subsection{Hong-Ou-Mandel two-photon interference}\label{Sup_subsec4}

\begin{figure}[hbt!]
 \centering
  \includegraphics[width=0.95\textwidth]{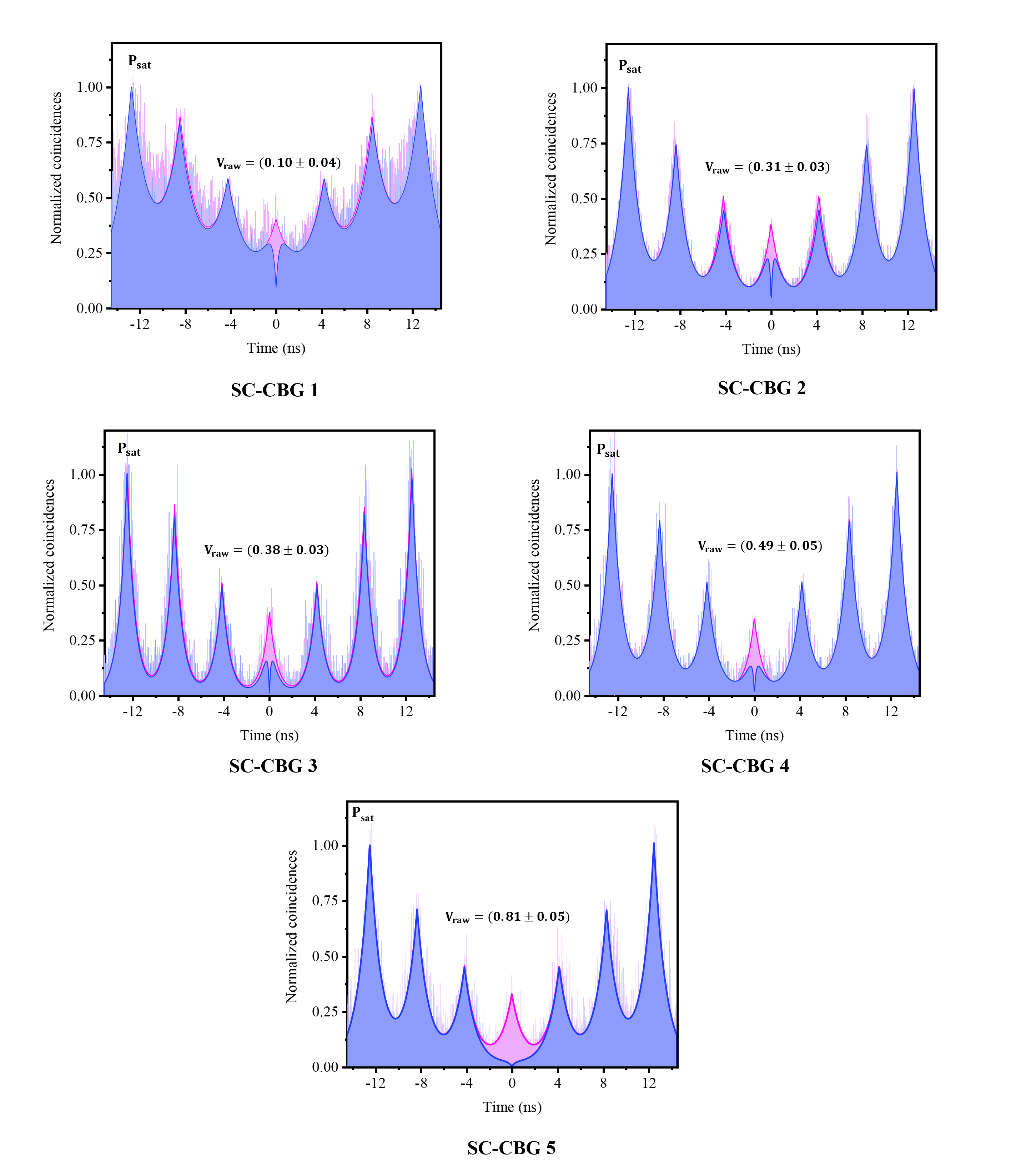}
  \caption{HOM TPI measurements for the five devices (SC-CBG 1-5) with different radial offsets, illustrating offset-dependent variations in photon indistinguishability. 
}\label{fig:S10}
\end{figure}
Fig. S \ref{fig:S10} shows the raw (and fitted) HOM TPI histograms for the five SCQD-CBG devices. The extracted HOM visibilities reveal a clear dependence on QD-resonator alignment, with higher indistinguishability for well-centered emitters and a systematic reduction as the radial offset increases. These data provide a direct experimental view of how emitter positioning affects TPI across devices.

\clearpage
\section{Modeling of charge noise amplitude}\label{Sup_sec6}
\begin{figure}
  \centering
  \includegraphics[width=.4\textwidth]{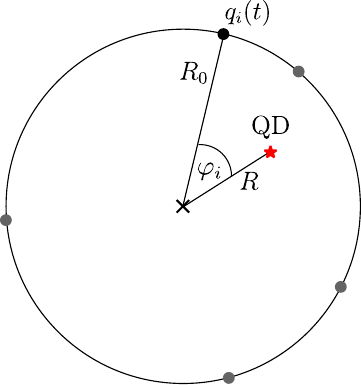}
  \caption{Sketch of the model underlying the calculation of the charge noise amplitude. The quantum dot with radial offset $R$ is displayed in red, while the charge fluctuators are randomly distributed on the etched surface of the CBG and displayed as gray dots. For one fluctuator, the definition of angle $\varphi_i$ is shown.}
  \label{fig:cn_model}
\end{figure}

To get an estimate of the charge noise amplitude, we assume that the most dominant contributions to the charge noise stem from surface states at the CBG boundary~\cite{Connors2019, Manna2020}. At each surface charge trap, a fluctuating charge $q_i(t)$ is located, inducing an electric field ${F}_i(\vec{r},t)$. The total electric field at the QD position $\vec{R}$ is then given by
\begin{align}
  F(\vec{R}, t) = \sum_i {F}_i(\vec{R}, t) = \sum_i \frac{1}{4 \pi \epsilon_0\epsilon_\mathrm{r}} \frac{q_i(t)}{|\vec{R} - \vec{r}_i|^2}.
\end{align}
This electric field will cause an energy shift of
\begin{align}
  \Delta E(\vec{R}, t) = \sum_i \frac{1}{4 \pi \epsilon_0\epsilon_\mathrm{r}} \frac{q_i(t)}{|\vec{R} - \vec{r}_i|^2}d_i
\end{align}
with $d_i$ being the projection of the dipole moment onto the direction of the electric field. This permanent dipole moment is caused by the average charge environment. The spectral noise amplitude at the QD location is given by
\begin{align}
  S(\omega) = \int \mathrm{d} t \mathrm{e}^{\mathrm{i} \omega t} \braket{\Delta E(\vec{R}, t) \Delta E(\vec{R}, 0)} \propto \sum_{ij} \int \mathrm{d} t\, \mathrm{e}^{\mathrm{i} \omega t} \braket{q_i(t) q_j(0)}  \frac{d_i d_j}{|\vec{R} - \vec{r}_i|^2|\vec{R} - \vec{r}_j|^2}.
\end{align}
The fluctuations at different sites are assumed to be independent, which leads to vanishing cross-correlations for $i \neq j$. We define $f_i = \int \mathrm{d} t \mathrm{e}^{\mathrm{i} \omega_0 t} d_i^2 \braket{q_i(t) q_i(0)}$ as the weighted noise spectrum of a single fluctuator at a fixed frequency $\omega_0$. The noise amplitude at this frequency is then given by
\begin{align}
  S := S(\omega_0) \propto \sum_{i} f_i(\omega)  \frac{1}{|\vec{R} - \vec{r}_i|^4} = \sum_i \frac{f_i(\omega)}{(R^2 + R_0^2 - 2RR_0 \cos \varphi_i)^2}.
\end{align}
Here, $\varphi_i$ is the relative angle as defined in Fig. S~\ref{fig:cn_model}, $R$ is the radial displacement of the QD, and $R_0$ is the radius of the CBG.
Introducing a fluctuator density $\rho(\varphi) = \sum_i f_i \delta (\varphi - \varphi_i)$, we can rewrite this as an integral:
\begin{align}
  S \propto \int_0^{2\pi} \frac{\mathrm{d} \varphi \, \rho(\varphi)}{{(R^2 + R_0^2 - 2RR_0 \cos \varphi)^2}} .
\end{align}
In general, the fluctuator density will depend on the sample and is unknown. To get an estimate for the average trend of the amplitude of the charge noise in a random sample, we assume it to be constant, since, on average, there is no angle preference. Then, we write the average noise amplitude as
\begin{align}
  S \propto  \int_0^{2\pi} \frac{\mathrm{d} \varphi}{{(R^2 + R_0^2 - 2RR_0 \cos \varphi)^2}}= 2\pi \frac{R_0^2 + R^2}{(R_0^2 - R^2)^3}.
\end{align}
For a single sample, this assumption is unjustified, but we use it here to get a general trend of the average charge noise amplitude as a function of the radial displacement. The dephasing rate $\gamma_\mathrm{deph}$ due to charge fluctuations will be proportional to the noise amplitude:
\begin{align}
  \gamma_\mathrm{deph}(R) = c \frac{R_0^2 + R^2}{(R_0^2 - R^2)^3}.
\end{align}
Together with the radiative recombination rate $\Gamma_\mathrm{rad} = T_1^{-1}$, we can calculate the Lorentzian linewidth as
\begin{align}
  \kappa = \Gamma_\mathrm{rad} + \gamma_\mathrm{deph}
\end{align}
and determine the proportionality constant $c$ by fitting to the measured linewidth. From this we get a proportionality factor of $c=\SI{0.220}{\nano \second^{-1} \micro \meter^{4}}$. The Hong-Ou-Mandel indistinguishability can be calculated using
\begin{align}
  \mathcal{I} = \frac{\int_0^\infty \mathrm{d} t \int_0^\infty \mathrm{d} \tau \left|C(t,\tau)\right|^2}{\int_0^\infty \mathrm{d} t \int_0^\infty \mathrm{d} \tau \, C(t,0) C(t+\tau,0)},
\end{align}
where $C(t,\tau)$ is the two-time correlation function of the optical transition. For a two-level system with Markovian dephasing, this correlation function is given by 
\begin{align}
  C(t, \tau) = \mathrm{e}^{-\Gamma_\mathrm{rad} t - (\Gamma_\mathrm{rad} + \gamma_\mathrm{deph}) \tau / 2},
\end{align}
resulting in an indistinguishability of $\mathcal{I} = \frac{\Gamma_\mathrm{rad}}{\Gamma_\mathrm{rad} + \gamma_\mathrm{deph}}$. For the computational results plotted in Fig.~5 in the main text, we chose a constant recombination rate $\Gamma_\mathrm{rad} = 1.25 \, \mathrm{ns}^{-1}$ to better showcase the general trend without being restricted to only five data points where the lifetime has been measured.
\begin{figure}
  \centering
  \includegraphics[width=.99\textwidth]{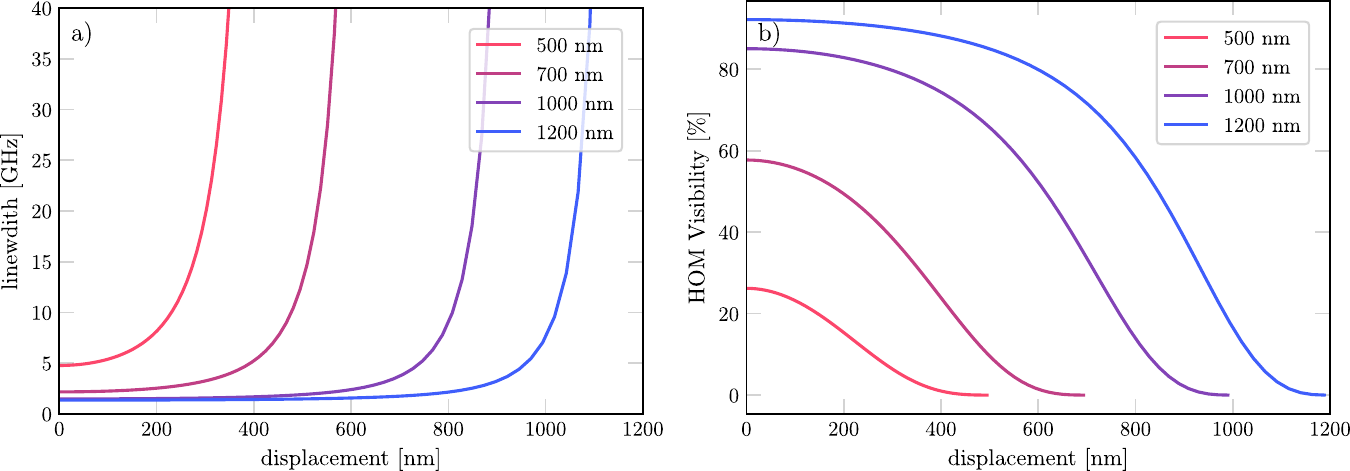}
  \caption{Theoretically calculated linewidth (a) and HOM visibility (b) as a function of the radial displacements for different mesa radii.}
  \label{fig:diff_r}
\end{figure}
Fig.~S~\ref{fig:diff_r} shows both the linewidth and HOM visibility calculated for different mesa radii $R_0$. As expected, a larger mesa diameter leads to a lower linewidth and higher visibility for equal alignment accuracies. Also, the saturation displacement, i.e., the displacement range in which better alignment only marginally increases the visibility, is larger for larger mesas. If one considers a mesa with radius $R_0 = 1200 \, \mathrm{nm}$, an improvement in alignment accuracy only slightly improves the HOM visibility. It needs to be pointed out that these results assume the same fitting constant $c$ (which includes the charge trap density on the surface) and radiative lifetime for all radii, which might vary with mesa diameter.

\bibliography{sn-bibliography}